\documentclass[reprint,aps,prd,superscriptaddress]{revtex4-2}
\usepackage{latexsym, amsmath, amssymb,amsthm,amsopn,amsfonts, graphicx, epstopdf, multirow}
\usepackage{color}
\usepackage{textcomp}
\usepackage{placeins}
\usepackage[dvipsnames]{xcolor}
\usepackage{adjustbox}
\usepackage{cancel}
\usepackage{orcidlink}
\usepackage{tabularx}
\usepackage{booktabs}
\usepackage{float}
\usepackage{siunitx}
\usepackage{graphicx}
\usepackage{subcaption}
\usepackage{verbatim}
\usepackage{hyperref}
\usepackage{nameref}
\usepackage{soul}
\usepackage{cleveref}

\definecolor{matblue}{rgb}{0 0.447 0.741}
\definecolor{matred}{rgb}{0.85 0.3250 0.098}

\hypersetup{linktoc=all}
\hypersetup{colorlinks = true, allcolors = blue}

\begin{document}
\title{BICEP/\textit{Keck} XXI: Constraints on early-Universe parity violation from multipole-dependent birefringence}

% To use this file: simply \input{author_list_<paper_id>.tex} in your main TeX file where you want the author list to go!
% generated for paper: Birefringence 2
% 2026-02-19 20:18:02
\collaboration{BICEP/\textit{Keck} Collaboration}
\author{P.~A.~R.~Ade}%
\affiliation{School of Physics and Astronomy, Cardiff University, Cardiff, CF24 3AA, United Kingdom}%
\author{Z.~Ahmed~\orcidlink{0000-0002-9957-448X}}%
\affiliation{Kavli Institute for Particle Astrophysics and Cosmology, Stanford University, Stanford, CA 94305, USA}%
\affiliation{SLAC National Accelerator Laboratory, Menlo Park, CA 94025, USA}%
\author{M.~Amiri~\orcidlink{0000-0001-6523-9029}}%
\affiliation{Department of Physics and Astronomy, University of British Columbia, Vancouver, British Columbia, V6T 1Z1, Canada}%
\author{D.~Barkats~\orcidlink{0000-0002-8971-1954}}%
\affiliation{Center for Astrophysics, Harvard \& Smithsonian, Cambridge, MA 01238, USA}%
\author{R.~Basu~Thakur~\orcidlink{0000-0002-3351-3078}}%
\affiliation{Department of Physics, California Institute of Technology, Pasadena, CA 91125, USA}%
\author{C.~A.~Bischoff~\orcidlink{0000-0001-9185-6514}}%
\affiliation{Department of Physics, University of Cincinnati, Cincinnati, OH 45221, USA}%
\author{D.~Beck~\orcidlink{0000-0003-0848-2756}}%
\email{dobeck@stanford.edu}%
\affiliation{Department of Physics, Stanford University, Stanford, CA 94305, USA}%
\author{J.~J.~Bock}%
\affiliation{Department of Physics, California Institute of Technology, Pasadena, CA 91125, USA}%
\affiliation{Jet Propulsion Laboratory, California Institute of Technology, Pasadena, CA 91109, USA}%
\author{H.~Boenish}%
\affiliation{Center for Astrophysics, Harvard \& Smithsonian, Cambridge, MA 01238, USA}%
\author{V.~Buza}%
\affiliation{Kavli Institute for Cosmological Physics, University of Chicago, Chicago, IL 60637, USA}%
\author{B.~Cantrall~\orcidlink{0000-0003-4541-7080}}%
\affiliation{Department of Physics, Stanford University, Stanford, CA 94305, USA}%
\affiliation{Kavli Institute for Particle Astrophysics and Cosmology, Stanford University, Stanford, CA 94305, USA}%
\author{J.~R.~Cheshire~IV~\orcidlink{0000-0002-1630-7854}}%
\affiliation{Department of Physics, California Institute of Technology, Pasadena, CA 91125, USA}%
\author{J.~Connors}%
\affiliation{National Institute of Standards and Technology, Boulder, CO 80305, USA}%
\author{J.~Cornelison~\orcidlink{0000-0002-2088-7345}}%
\affiliation{High-Energy Physics Division, Argonne National Laboratory, Lemont, IL, 60439, USA}%
\author{M.~Crumrine}%
\affiliation{School of Physics and Astronomy, University of Minnesota, Minneapolis, MN 55455, USA}%
\author{A.~J.~Cukierman}%
\affiliation{Department of Physics, California Institute of Technology, Pasadena, CA 91125, USA}%
\author{E.~Denison}%
\affiliation{National Institute of Standards and Technology, Boulder, CO 80305, USA}%
\author{L.~Duband}%
\affiliation{Service des Basses Temp\'eratures, Commissariat \`a l'\'Energie Atomique, 38054 Grenoble, France}%
\author{M.~Echter}%
\affiliation{Center for Astrophysics, Harvard \& Smithsonian, Cambridge, MA 01238, USA}%
\author{M.~Eiben~\orcidlink{0009-0007-6718-1730}}%
\affiliation{Faculty of Physical Sciences, University of Iceland, 102 Reykjavík, Iceland}%
\author{B.~D.~Elwood~\orcidlink{0000-0003-4117-6822}}%
\affiliation{Center for Astrophysics, Harvard \& Smithsonian, Cambridge, MA 01238, USA}%
\affiliation{Department of Physics, Harvard University, Cambridge, MA 02138, USA}%
\author{S.~Fatigoni~\orcidlink{0000-0002-3790-7314}}%
\affiliation{Department of Physics, California Institute of Technology, Pasadena, CA 91125, USA}%
\author{J.~P.~Filippini~\orcidlink{0000-0001-8217-6832}}%
\affiliation{Department of Physics, University of Illinois at Urbana-Champaign, Urbana, IL 61801, USA}%
\author{A.~Fortes}%
\affiliation{Department of Physics, Stanford University, Stanford, CA 94305, USA}%
\author{M.~Gao}%
\affiliation{Department of Physics, California Institute of Technology, Pasadena, CA 91125, USA}%
\author{C.~Giannakopoulos}%
\affiliation{Department of Physics, University of Cincinnati, Cincinnati, OH 45221, USA}%
\author{N.~Goeckner-Wald}%
\affiliation{Department of Physics, Stanford University, Stanford, CA 94305, USA}%
\author{D.~C.~Goldfinger~\orcidlink{0000-0001-5268-8423}}%
\affiliation{Department of Physics, Stanford University, Stanford, CA 94305, USA}%
\author{S.~Gratton}%
\affiliation{Centre for Theoretical Cosmology, DAMTP, University of Cambridge, Cambridge CB3 0WA, UK}%
\affiliation{Kavli Institute for Cosmology Cambridge, Cambridge CB3 0HA, UK}%
\author{J.~A.~Grayson}%
\affiliation{Department of Physics, Stanford University, Stanford, CA 94305, USA}%
\author{A.~Greathouse~\orcidlink{0009-0003-6999-0129}}%
\affiliation{Department of Physics, California Institute of Technology, Pasadena, CA 91125, USA}%
\author{P.~K.~Grimes~\orcidlink{0000-0001-9292-6297}}%
\affiliation{Center for Astrophysics, Harvard \& Smithsonian, Cambridge, MA 01238, USA}%
\author{G.~Hall}%
\affiliation{Minnesota Institute for Astrophysics, University of Minnesota, Minneapolis, MN 55455, USA}%
\affiliation{Department of Physics, Stanford University, Stanford, CA 94305, USA}%
\author{G.~Halal~\orcidlink{0000-0003-2221-3018}}%
\affiliation{Department of Physics, Stanford University, Stanford, CA 94305, USA}%
\author{M.~Halpern}%
\affiliation{Department of Physics and Astronomy, University of British Columbia, Vancouver, British Columbia, V6T 1Z1, Canada}%
\author{E.~Hand}%
\affiliation{Department of Physics, University of Cincinnati, Cincinnati, OH 45221, USA}%
\author{S.~A.~Harrison}%
\affiliation{Center for Astrophysics, Harvard \& Smithsonian, Cambridge, MA 01238, USA}%
\author{S.~Henderson}%
\affiliation{Kavli Institute for Particle Astrophysics and Cosmology, Stanford University, Stanford, CA 94305, USA}%
\affiliation{SLAC National Accelerator Laboratory, Menlo Park, CA 94025, USA}%
\author{T.~D.~Hoang~\orcidlink{0000-0002-3437-5228}}%
\affiliation{School of Physics and Astronomy, University of Minnesota, Minneapolis, MN 55455, USA}%
\author{J.~Hubmayr}%
\affiliation{National Institute of Standards and Technology, Boulder, CO 80305, USA}%
\author{H.~Hui~\orcidlink{0000-0001-5812-1903}}%
\affiliation{Department of Physics, California Institute of Technology, Pasadena, CA 91125, USA}%
\author{K.~D.~Irwin}%
\affiliation{Department of Physics, Stanford University, Stanford, CA 94305, USA}%
\author{J.~H.~Kang~\orcidlink{0000-0002-3470-2954}}%
\affiliation{Department of Physics, California Institute of Technology, Pasadena, CA 91125, USA}%
\author{K.~S.~Karkare~\orcidlink{0000-0002-5215-6993}}%
\affiliation{Department of Physics, Boston University, Boston, MA 02215, USA}%
\author{S.~Kefeli}%
\affiliation{Department of Physics, California Institute of Technology, Pasadena, CA 91125, USA}%
\author{J.~M.~Kovac~\orcidlink{0009-0003-5432-7180}}%
\affiliation{Center for Astrophysics, Harvard \& Smithsonian, Cambridge, MA 01238, USA}%
\affiliation{Department of Physics, Harvard University, Cambridge, MA 02138, USA}%
\author{C.~Kuo}%
\affiliation{Department of Physics, Stanford University, Stanford, CA 94305, USA}%
\author{K.~Lasko~\orcidlink{0000-0002-4540-1495}}%
\affiliation{School of Physics and Astronomy, University of Minnesota, Minneapolis, MN 55455, USA}%
\affiliation{Minnesota Institute for Astrophysics, University of Minnesota, Minneapolis, MN 55455, USA}%
\author{K.~Lau~\orcidlink{0000-0002-6445-2407}}%
\affiliation{Department of Physics, California Institute of Technology, Pasadena, CA 91125, USA}%
\author{M.~Lautzenhiser}%
\affiliation{Department of Physics, University of Cincinnati, Cincinnati, OH 45221, USA}%
\author{A.~Lennox}%
\affiliation{Department of Physics, University of Illinois at Urbana-Champaign, Urbana, IL 61801, USA}%
\author{T.~Liu~\orcidlink{0000-0001-5677-5188}}%
\email{tongtianliu@stanford.edu}%
\affiliation{Department of Physics, Stanford University, Stanford, CA 94305, USA}%
\author{S.~C.~Mackey~\orcidlink{0000-0002-1414-7236}}%
\affiliation{Kavli Institute for Cosmological Physics, University of Chicago, Chicago, IL 60637, USA}%
\affiliation{Department of Physics, University of Chicago, Chicago, IL 60637, USA}%
\author{N.~Maher}%
\affiliation{School of Physics and Astronomy, University of Minnesota, Minneapolis, MN 55455, USA}%
\author{K.~G.~Megerian}%
\affiliation{Jet Propulsion Laboratory, California Institute of Technology, Pasadena, CA 91109, USA}%
\author{L.~Minutolo}%
\affiliation{Department of Physics, California Institute of Technology, Pasadena, CA 91125, USA}%
\author{L.~Moncelsi~\orcidlink{0000-0002-4242-3015}}%
\affiliation{Department of Physics, California Institute of Technology, Pasadena, CA 91125, USA}%
\author{Y.~Nakato}%
\affiliation{Department of Physics, Stanford University, Stanford, CA 94305, USA}%
\author{H.~T.~Nguyen}%
\affiliation{Department of Physics, California Institute of Technology, Pasadena, CA 91125, USA}%
\affiliation{Jet Propulsion Laboratory, California Institute of Technology, Pasadena, CA 91109, USA}%
\author{R.~O’Brient}%
\affiliation{Department of Physics, California Institute of Technology, Pasadena, CA 91125, USA}%
\affiliation{Jet Propulsion Laboratory, California Institute of Technology, Pasadena, CA 91109, USA}%
\author{S.~N.~Paine}%
\affiliation{Center for Astrophysics, Harvard \& Smithsonian, Cambridge, MA 01238, USA}%
\author{A.~Patel}%
\affiliation{Department of Physics, California Institute of Technology, Pasadena, CA 91125, USA}%
\author{M.~A.~Petroff~\orcidlink{0000-0002-4436-4215}}%
\affiliation{Center for Astrophysics, Harvard \& Smithsonian, Cambridge, MA 01238, USA}%
\author{A.~R.~Polish~\orcidlink{0000-0002-7822-6179}}%
\affiliation{Center for Astrophysics, Harvard \& Smithsonian, Cambridge, MA 01238, USA}%
\affiliation{Department of Physics, Harvard University, Cambridge, MA 02138, USA}%
\author{T.~Prouve}%
\affiliation{Service des Basses Temp\'eratures, Commissariat \`a l'\'Energie Atomique, 38054 Grenoble, France}%
\author{C.~Pryke~\orcidlink{0000-0003-3983-6668}}%
\affiliation{School of Physics and Astronomy, University of Minnesota, Minneapolis, MN 55455, USA}%
\author{C.~D.~Reintsema}%
\affiliation{National Institute of Standards and Technology, Boulder, CO 80305, USA}%
\author{S.~Richter}%
\affiliation{Center for Astrophysics, Harvard \& Smithsonian, Cambridge, MA 01238, USA}%
\author{T.~Romand}%
\affiliation{Department of Physics, California Institute of Technology, Pasadena, CA 91125, USA}%
\author{M.~Salatino}%
\affiliation{Department of Physics, Stanford University, Stanford, CA 94305, USA}%
\author{A.~Schillaci}%
\affiliation{Department of Physics, California Institute of Technology, Pasadena, CA 91125, USA}%
\author{B.~Schmitt}%
\affiliation{Center for Astrophysics, Harvard \& Smithsonian, Cambridge, MA 01238, USA}%
\author{R.~Schwartz}%
\affiliation{School of Physics and Astronomy, University of Minnesota, Minneapolis, MN 55455, USA}%
\author{C.~D.~Sheehy}%
\affiliation{School of Physics and Astronomy, University of Minnesota, Minneapolis, MN 55455, USA}%
\author{B.~Singari~\orcidlink{0000-0001-7387-0881}}%
\affiliation{School of Physics and Astronomy, University of Minnesota, Minneapolis, MN 55455, USA}%
\affiliation{Minnesota Institute for Astrophysics, University of Minnesota, Minneapolis, MN 55455, USA}%
\author{A.~Soliman}%
\affiliation{Department of Physics, California Institute of Technology, Pasadena, CA 91125, USA}%
\affiliation{Jet Propulsion Laboratory, California Institute of Technology, Pasadena, CA 91109, USA}%
\author{T.~St.~Germaine}%
\affiliation{Center for Astrophysics, Harvard \& Smithsonian, Cambridge, MA 01238, USA}%
\author{A.~Steiger~\orcidlink{0000-0003-0260-605X}}%
\affiliation{Department of Physics, California Institute of Technology, Pasadena, CA 91125, USA}%
\author{B.~Steinbach}%
\affiliation{Department of Physics, California Institute of Technology, Pasadena, CA 91125, USA}%
\author{R.~Sudiwala}%
\affiliation{School of Physics and Astronomy, Cardiff University, Cardiff, CF24 3AA, United Kingdom}%
\author{G.~Teply}%
\affiliation{Department of Physics, California Institute of Technology, Pasadena, CA 91125, USA}%
\author{K.~L.~Thompson}%
\affiliation{Kavli Institute for Particle Astrophysics and Cosmology, Stanford University, Stanford, CA 94305, USA}%
\affiliation{Department of Physics, Stanford University, Stanford, CA 94305, USA}%
\author{C.~Tucker~\orcidlink{0000-0002-1851-3918}}%
\affiliation{School of Physics and Astronomy, Cardiff University, Cardiff, CF24 3AA, United Kingdom}%
\author{A.~D.~Turner}%
\affiliation{Jet Propulsion Laboratory, California Institute of Technology, Pasadena, CA 91109, USA}%
\author{C.~Verg\`{e}s~\orcidlink{0000-0002-3942-1609}}%
\affiliation{Physics Division, Lawrence Berkeley National Laboratory, Berkeley, CA 94720, USA}%
\author{A.~G.~Vieregg}%
\affiliation{Kavli Institute for Cosmological Physics, University of Chicago, Chicago, IL 60637, USA}%
\affiliation{Department of Physics, University of Chicago, Chicago, IL 60637, USA}%
\author{A.~Wandui~\orcidlink{0000-0002-8232-7343}}%
\affiliation{Department of Physics, California Institute of Technology, Pasadena, CA 91125, USA}%
\author{A.~C.~Weber}%
\affiliation{Jet Propulsion Laboratory, California Institute of Technology, Pasadena, CA 91109, USA}%
\author{J.~Willmert~\orcidlink{0000-0002-6452-4693}}%
\affiliation{School of Physics and Astronomy, University of Minnesota, Minneapolis, MN 55455, USA}%
\author{C.~L.~Wong}%
\affiliation{Center for Astrophysics, Harvard \& Smithsonian, Cambridge, MA 01238, USA}%
\affiliation{Department of Physics, Harvard University, Cambridge, MA 02138, USA}%
\author{W.~L.~K.~Wu~\orcidlink{0000-0001-5411-6920}}%
\affiliation{Kavli Institute for Particle Astrophysics and Cosmology, Stanford University, Stanford, CA 94305, USA}%
\affiliation{SLAC National Accelerator Laboratory, Menlo Park, CA 94025, USA}%
\author{H.~Yang}%
\affiliation{Department of Physics, Stanford University, Stanford, CA 94305, USA}%
\author{C.~Yu~\orcidlink{0000-0002-8542-232X}}%
\affiliation{Kavli Institute for Cosmological Physics, University of Chicago, Chicago, IL 60637, USA}%
\affiliation{High-Energy Physics Division, Argonne National Laboratory, Lemont, IL, 60439, USA}%
\author{L.~Zheng~\orcidlink{0000-0001-6924-9072}}%
\affiliation{Center for Astrophysics, Harvard \& Smithsonian, Cambridge, MA 01238, USA}%
\author{C.~Zhang~\orcidlink{0000-0001-8288-5823}}%
\affiliation{Kavli Institute for Particle Astrophysics and Cosmology, Stanford University, Stanford, CA 94305, USA}%
\author{S.~Zhang}%
\affiliation{Department of Physics, California Institute of Technology, Pasadena, CA 91125, USA}

\begin{abstract}
We present the first constraints on multipole-dependent cosmic birefringence using CMB polarization data from the BK18 dataset, which combines observations from BICEP2, \textit{Keck} Array, and BICEP3 at frequencies of 95, 150, and 220 GHz. Photon coupling to an axionlike field leads to the rotation of CMB polarization, inducing nonzero $EB$ cross-correlations. We show that a multipole-dependent rotation $\beta(\ell)$ imprints a distinct signature in the polarization spectra that can be constrained. Specifically, we consider an early dark energy (EDE) scenario in which a pseudoscalar field couples to photons through a Chern–Simons interaction, generating a polarization rotation with multipole dependence. We introduce a phenomenological $\beta(\ell)$ as a step function, obtaining constraints on the step function size consistent with zero, with uncertainties less than $\sim$0.15$^\circ$ (68\% CL). In addition, using multifrequency $EE$, $BB$, and $EB$ cross spectra, along with robust BICEP/\textit{Keck} foreground treatment and likelihood framework, we derive constraints on the axion–photon coupling amplitude $g$ for several choices of EDE parameters. For the baseline best-fit value $f_{\mathrm{EDE}} = 0.087$ from the \textit{Planck} 2018 analysis, we obtain $g = 0.11 \pm 0.37$ (68\% CL), consistent with previous limits.  

%All analyses are validated using comprehensive simulation ensembles, including signal-only and signal-plus-noise realizations processed through the full BICEP/\textit{Keck} analysis pipeline. These results demonstrate the ability of current CMB polarization data to probe the presence of a time-varying axionlike field at recombination and motivate future studies incorporating more general multipole-dependent forms of $\beta(\ell)$.
\end{abstract}

\maketitle

\section{Introduction}
\label{sec:introduction}

Cosmic birefringence refers to a rotation of the linear polarization plane of electromagnetic radiation as it propagates through space~\cite{Finelli2009,MinamiKomatsu2020,Xia2010,DiegoPalazuelos2022,Eskilt2022, Liu2006}. Such a rotation may arise from a coupling of photons to a pseudoscalar field, as predicted in various extensions of the standard model~\cite{Finelli2009,Carroll1990,Greco2024,Fujita2021}. A detection of this effect would provide direct evidence for new physics, such as axionlike particles, which have been proposed as dark matter candidates~\cite{Fujita2021,Nakagawa2023,Harari1992}.

The cosmic microwave background (CMB) is a particularly sensitive probe of cosmic birefringence~\cite{MinamiKomatsu2020,DiegoPalazuelos2022, Eskilt2022}. The linear polarization of the CMB can be decomposed into even-parity $E$ modes and odd-parity $B$ modes~\cite{Kamionkowski1997, Zaldarriaga1997,BICEP2Keck2015}. There is no expected correlation between these modes, and the $EB$ cross spectrum should vanish~\cite{Lue1999, Kamionkowski2009}. A statistically significant detection of a nonzero $EB$ correlation would thus suggest the presence of axionlike fields through Chern-Simons type interactions~\cite{Wu2009,MinamiKomatsu2020}.

Measuring a global rotation is experimentally challenging because it requires knowing the instrumental polarization orientation of each detector with sufficient precision. A uniform per-band effective rotation by an angle $\alpha$, which results from the aggregate instrumental response of a given frequency band, is degenerate with a global birefringence angle $\beta$, such that the observed $EB$ spectrum reflects the sum $\alpha + \beta$~\cite{MinamiKomatsu2020,DiegoPalazuelos2022}. In this isotropic scenario, the induced $EB$ power spectrum shape is largely proportional to the $EE$ spectrum.

 Several strategies have been proposed to break this degeneracy. The BICEP/\textit{Keck} (BK) collaboration has initiated a dedicated polarization angle calibration campaign~\cite{BICEPKeck2025,Cornelison2022} to enable forward-modeling constraints on $\beta$ using known detector angles to measure global birefringence.  Alternatively, a new dynamic field can manifest in anisotropic polarization rotation~\cite{Bianchini2020, Contreras2017, Gruppuso_2020, Zagatti_2024, Bortolami_2022} or time-evolving cosmic birefringence~\cite{Ferguson2022}, which possess distinct signatures that do not require absolute angle calibration to measure. These effects have also been studied in the BK data~\cite{BICEPKeckIX, BICEPKeckXII, BICEPKeckXIV, BICEPKeckXVII}. 
Another proposed method is to use polarized Galactic dust emission as a calibrator, modeling its intrinsic $EB$ dust as proportional to the $EE$ dust~\cite{Eskilt2022}; however, this assumption has been questioned and demonstrated to be at least partially false~\cite{Cukierman2023}.

In this paper, we investigate the multipole dependence ($\ell$ dependence) of the rotation angle in BICEP/\textit{Keck} data. If the pseudoscalar field evolves during recombination, the epoch when CMB polarization was generated, the redshift-dependent rotation will lead to such $\ell$ dependence~\cite{Eskilt2023,Murai2023}. Since $EE$ modes are generated at different times throughout recombination, a time-varying $\beta(z)$ results in an $EB$ spectrum that encodes the evolving phase structure of the $EE$ peaks. In particular, as the sound horizon changes with redshift, the peaks in the rotated $EB$ spectrum reflect a shifted and temporally smeared version of the original $EE$ peaks~\cite{Bernal2016}. We parametrize this phenomenologically by a multipole-dependent rotation angle function $\beta(\ell)$. Measuring such $\ell$ dependence provides a concrete method to distinguish cosmic birefringence from a simple detector rotation~\cite{Sherwin2021, Ballardini_2024}.

An example of motivated models predicting $\ell$ dependent birefringence is early dark energy (EDE)~\cite{Karwal2016}. EDE posits a dynamical scalar field that becomes non-negligible around the epoch of matter-radiation equality. The model is characterized by three primary parameters: the maximum fractional energy density $f_{\mathrm{EDE}}$, the critical redshift $z_c$ at which this density is peaked, and the initial field displacement $\theta_i$~\cite{Herold2022,Smith2021,Reeves2023}. The field quickly diminishes during recombination from $z\approx1200$ to $z\approx900$, leaving minimal impact at late times, and modifies the expansion history in a way that can raise the value of $H_0$ inferred from the CMB, potentially alleviating the Hubble tension~\cite{Herold2022,HeroldThesis2023,Poulin2023}.

If the EDE field is a pseudoscalar axionlike particle coupled to electromagnetism via a Chern-Simons interaction, it induces a rotation of CMB polarization. The Lagrangian includes a term that gives rise to this effect:
\begin{equation}\label{eq:lagrange}
\mathcal{L} \supset -\frac{1}{4} g \phi F_{\mu\nu} \tilde{F}^{\mu\nu},
\end{equation}
where $g$ is the axion-photon coupling constant, $\phi$ is the EDE field, and $\tilde{F}^{\mu\nu}$ is the dual electromagnetic field tensor~\cite{Finelli2009,Nakagawa2023}.

\begin{figure*}[t]
    \centering
    \begin{subfigure}[t]{0.48\textwidth}
        \centering
        \includegraphics[width=\textwidth]{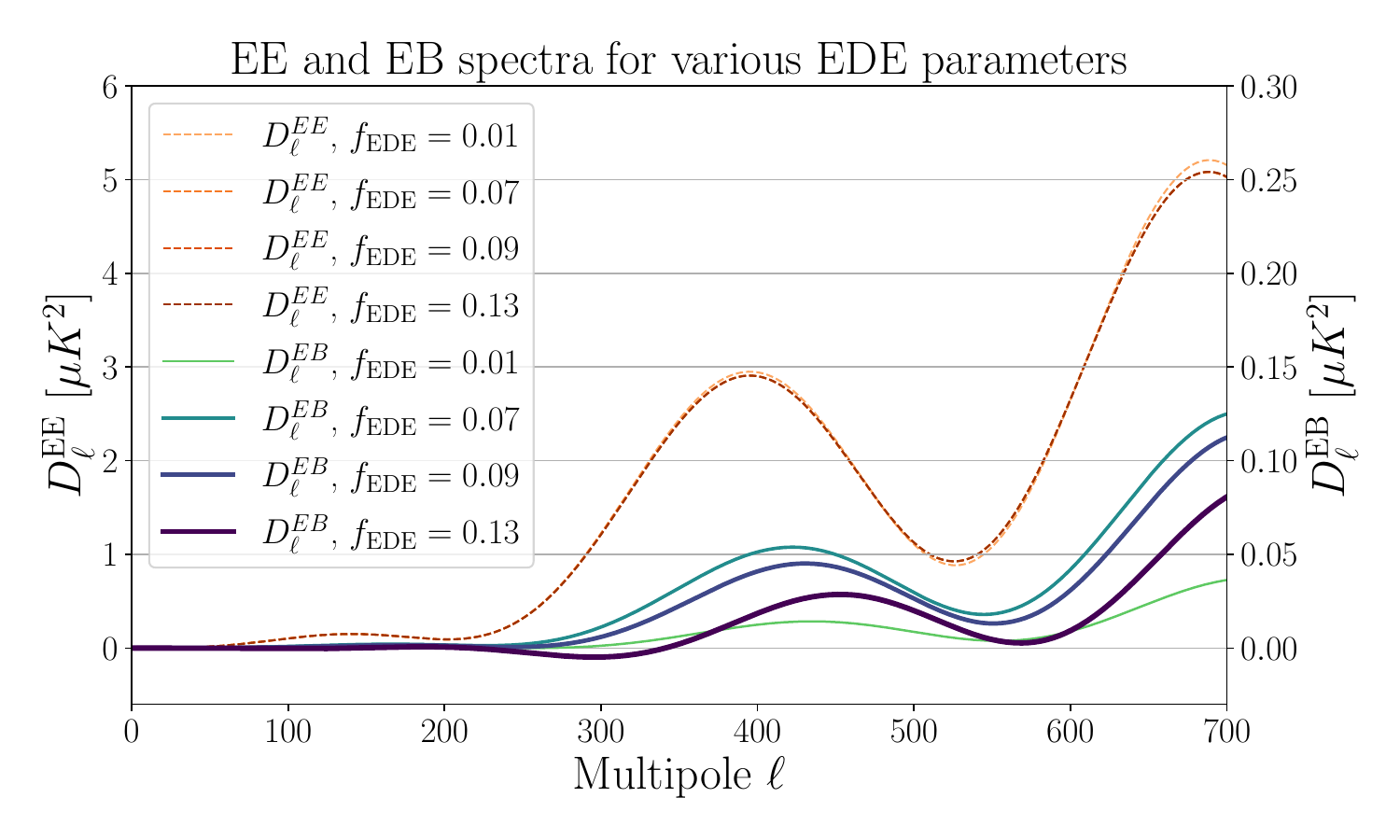}
    \end{subfigure}
    \hfill
    \begin{subfigure}[t]{0.48\textwidth}
        \centering
        \includegraphics[width=\textwidth]{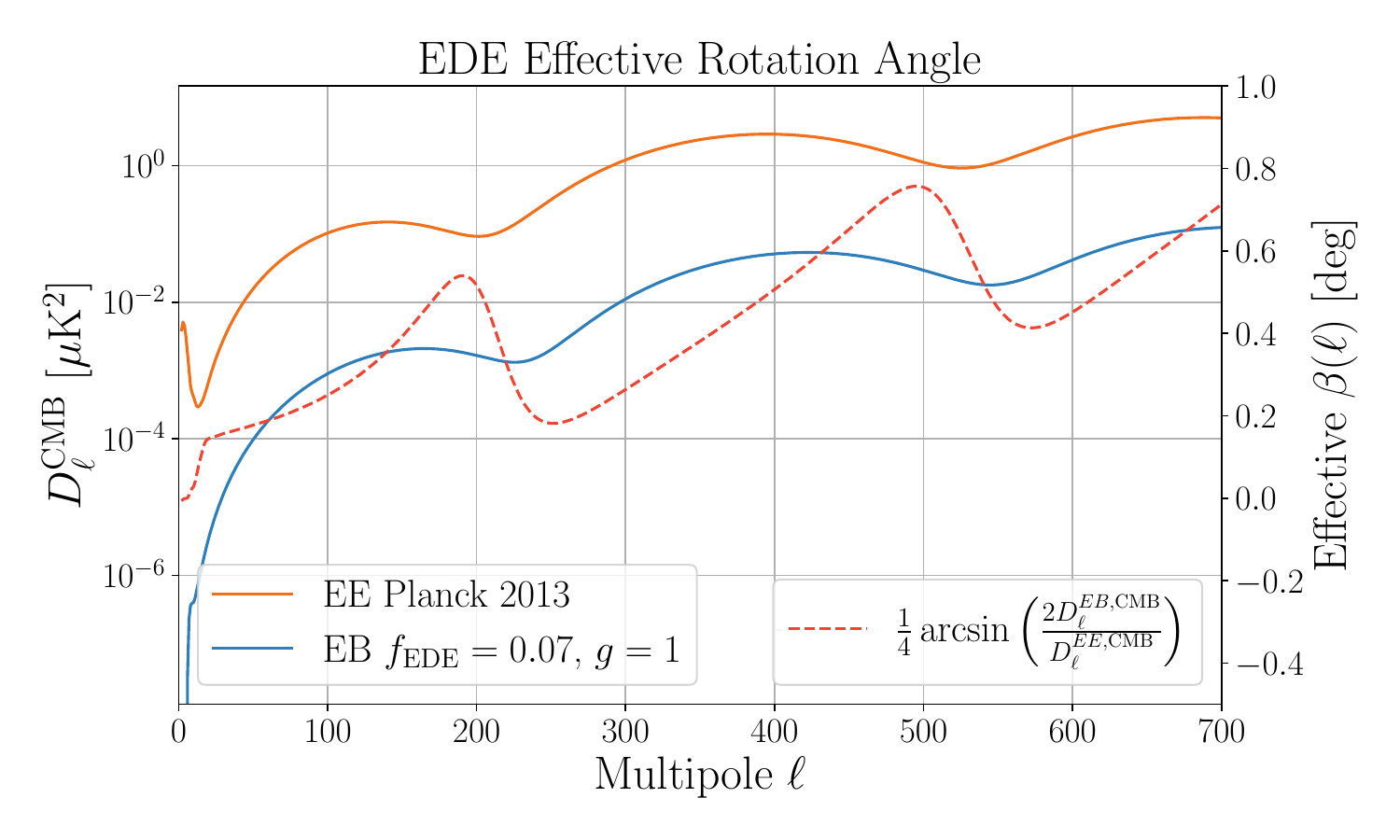}
    \end{subfigure}
    \caption{\textbf{Multipole dependence in early dark energy.} (Left) $EB$ and $EE$ spectra generated by early dark energy models with different values of $f_{\mathrm{EDE}}$. As $f_{\mathrm{EDE}}$ increases, the $EB$ spectrum (blue) becomes more pronounced and deviates further from a constant rotation model, while the $EE$ spectrum (orange) stays roughly the same. $f_{\mathrm{EDE}}$ is the primary driver of these shifts~\cite{Karwal2016,Hill2022, Herold2022}, but other parameters (both EDE and standard cosmological parameters) also vary according to best fit values in Table~\ref{tab:ede_bestfits_reordered}. Note that $EE$ spectra and $EB$ spectra are on different scales for visual comparison. (Right) Effective rotation angle (red) approximated by $\beta(\ell) = \frac{1}{4}\arcsin(2D_\ell^{EB} / D_\ell^{EE})$, where $EB$ (blue) is derived from an EDE model with $f_{\text{EDE}}=0.07$, and $EE$ (orange) is the best fit curve from \textit{Planck} 2013. Constant-angle models are insufficient for fitting to $\beta(\ell)$.}
    \label{fig:ede_combined}
\end{figure*}

The time evolution of $\phi$ during recombination generates the aforementioned redshift-dependent birefringence angle $\beta(z)$, which leads to an $EB$ spectrum with nontrivial $\ell$ dependence~\cite{Eskilt2023,Murai2023,Nakatsuka2022, Lee_2014_DarkEnergy}. This dependence varies with specific EDE parameters $(f_{\mathrm{EDE}}, z_c, \theta_i)$, as shown in Fig.~\ref{fig:ede_combined} (left). As the EDE fraction $f_{\mathrm{EDE}}$ increases, the amplitude of the $EB$ spectrum grows and its peak structure becomes increasingly inconsistent with the shape expected from a constant $\beta$. Notably, the corresponding $EE$ spectra remain largely unchanged, indicating that the $EB$ features cannot be adequately captured by a constant rotation angle alone.
A 2023 analysis by Eskilt \textit{et al.}~\cite{Eskilt2023} searched for this effect using \textit{Planck} polarization data. By comparing the model-predicted $EB$ spectrum to the observed stacked power spectrum, they found \(g = 0.04 \pm 0.16\) (68\% confidence). While their data did not favor the specific shape predicted by the model, their work demonstrated that the distinct shape of $EB$ can break the $\alpha$-$\beta$ degeneracy without requiring absolute calibration or assumptions about Galactic dust~\cite{Eskilt2023,DiegoPalazuelos2022}.

To quantify the difference in shape, we can define an effective rotation angle:
\begin{equation}
    \beta(\ell) = \frac{1}{4}\arcsin\left( \frac{2D_\ell^{EB}}{D_\ell^{EE}} \right),
\end{equation}
which approximates the apparent rotation angle at each $\ell$ under the assumption that $EB$ arises entirely from rotated $EE$. This function, shown in Fig.~\ref{fig:ede_combined} (right) for a representative EDE model, displays the multipole dependence of the rotation angle that we seek to constrain.

In this work, we extend the EDE analysis using data from the BICEP/\textit{Keck} Array~\cite{BICEPKeck2021b,BICEPKeck2025}. While BICEP/\textit{Keck} probes a smaller multipole range ($\ell \lesssim 520$) than \textit{Planck}, its high sensitivity, choice of a clean sky patch, and multifrequency coverage enable robust measurements of the $EB$ spectrum with low noise and foreground contamination~\cite{BICEP2Keck2015,Planck2015PMF}.

We present complementary analyses:
\begin{enumerate}
    \item \textbf{Model-independent search for multipole-dependent birefringence} (Sec.\ref{sec:angle_diff}): A novel approach modeling $\beta(\ell)$ as a step function, measuring $\Delta\beta = \beta_{\mathrm{low}\text{-}\ell} - \beta_{\mathrm{high}\text{-}\ell}$ as a signature of multipole-dependent physics independent of any specific EDE model.
    \item \textbf{EDE model fit} (Sec.~\ref{sec:ede}): A direct search of EDE-predicted polarization spectra incorporating multicomponent foreground analysis for representative EDE parameters.
\end{enumerate}

Our multipole-dependent approach generalizes previous work by decoupling the analysis from specific EDE models, offering direct sensitivity to a broader range of multipole-dependent physics.   

\section{Data Model and Likelihood}
\label{sec:isotropic_method}
In our analysis, we model the effects of gravitational lensing, isotropic cosmic birefringence, EDE, polarized foregrounds, and effective per-band polarization rotation angles on CMB polarization spectra.

The lensed CMB $EE$ and $BB$ power spectra are computed using \texttt{CAMB}~\cite{Lewis2000} with Planck 2013 cosmological parameters and no primordial tensor modes~\cite{Planck2013params}.  The $BB$ spectrum is scaled by an amplitude parameter $A_{\mathrm{lens}}$ to account for uncertainties in the lensing signal:
\begin{equation}\label{eq:alens}
    D_\ell^{BB,\mathrm{CMB}} = A_{\mathrm{lens}} \cdot D_\ell^{BB,\mathrm{CAMB}}.
\end{equation}

Typically, setting $A_{\mathrm{lens}} = 1$ is sufficient because the $BB$ spectrum has 2 to 3 orders of magnitude less impact than $EE$ in constraining rotation from the $EB$ spectrum. However, we allow it to vary when simultaneously fitting the full $EE$, $BB$, and $EB$ spectra alongside foreground components.
\begin{comment}

\begin{figure}[h]
    \centering
    \includegraphics[width=0.97\linewidth]{figures/main_plots/cls.pdf}
    \caption{\textbf{Visualizing input polarization spectra}. \(BB\) curves are generated using Planck 2013 parameters, and \(EB\) curves are generated from Planck best-fit EDE parameters at \(f_{\text{EDE}} = 0.07\). The \(BB\) spectrum is scaled by various values of \(A_{\mathrm{lens}}\), and the \(EB\) spectrum by different values of \(g\).}
    \label{fig:sample_cls}
\end{figure}
\subsection{Modeling Multipole-Dependent Birefringence and Early Dark Energy}
\label{app:init_spectra}
\end{comment}
The CMB $EB$ spectrum is modeled as a parity-violating contribution from EDE, calculated with a modified version of \texttt{CLASS}~\cite{Blas2011} developed by Murai \textit{et al.}~\cite{Murai2023}:
\begin{equation}
    D_\ell^{EB,\mathrm{CMB}} = g \cdot D_\ell^{EB,\mathrm{EDE}}, \label{eq:ede_eb}
\end{equation}
where $g$ from Eq.~\eqref{eq:lagrange} parametrizes the amplitude of the EDE contribution to the $EB$ spectrum, and $D_\ell^{EB,\mathrm{EDE}}$ is the template power spectrum derived from specific EDE parameters. The parameter $g$ has units of $M_{\mathrm{pl}}^{-1}$ and is fixed to zero unless explicitly fitting for an EDE signal. %Figure~\ref{fig:sample_cls} illustrates examples of the initial $EE$, $BB$, and $EB$ spectra for various values of $g$ and $A_{\mathrm{lens}}$.

Throughout this work, we denote $\tilde{D}_\ell$ as rotated CMB spectra, $\hat{D}_\ell$ as spectra with foregrounds included, and \(\bar{D}_\ell\) as spectra after applying per-band rotation angles.

A uniform rotation angle $\beta_{\mathrm{CMB}}$, representing isotropic cosmic birefringence, is applied to the lensed CMB spectra to obtain the following~\cite{Feng2005, Cai2022, Kamionkowski2009}:

\begin{figure*}[ht]
    \centering
    \includegraphics[width=0.95\linewidth]{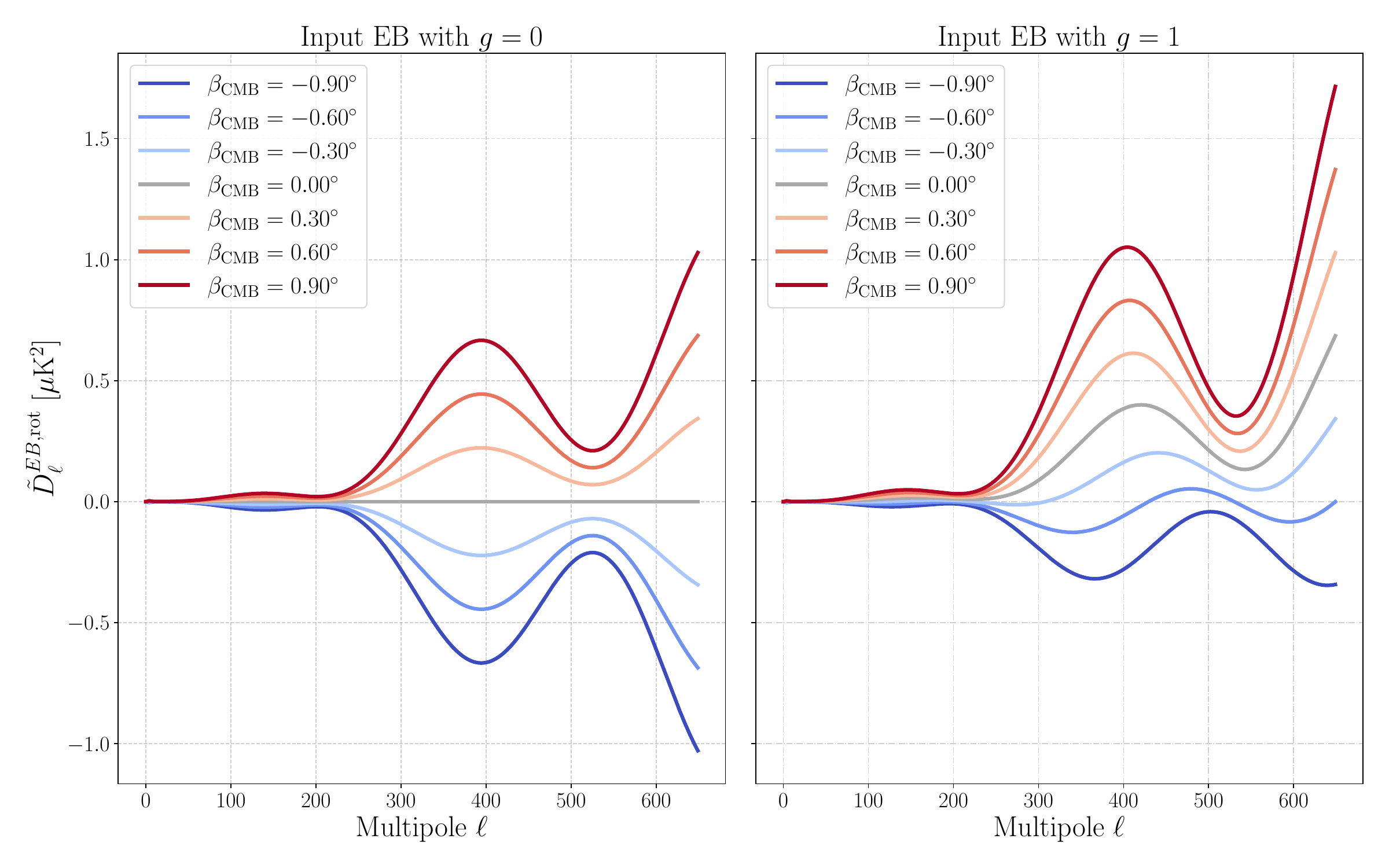}
    \caption{\textbf{Comparison of simulated CMB spectra under isotropic vs. multipole-dependent rotation.} (Left) Standard $\Lambda$CDM spectra rotated by different constant angles, $\beta_{\mathrm{CMB}}$ (isotropic birefringence only, $g=0$). (Right) Spectra including an early dark energy contribution ($g=1$) rotated by different $\beta_{\mathrm{CMB}}$. The EDE signal introduces distinct features in the $EB$ spectrum that differ significantly from the isotropic case.}
    \label{fig:sample_rotate}
\end{figure*}

\begin{widetext}\label{eq:rot_uni}
\begin{align}
    \tilde{D}_\ell^{EE,\mathrm{rot}} &= D_\ell^{EE,\mathrm{CMB}} \cos^2(2\beta_{\mathrm{CMB}}) + D_\ell^{BB,\mathrm{CMB}} \sin^2(2\beta_{\mathrm{CMB}}) - D_\ell^{EB,\mathrm{CMB}} \sin(4\beta_{\mathrm{CMB}}), \label{eq:rot_ee} \\
    \tilde{D}_\ell^{BB,\mathrm{rot}} &= D_\ell^{EE,\mathrm{CMB}} \sin^2(2\beta_{\mathrm{CMB}}) + D_\ell^{BB,\mathrm{CMB}} \cos^2(2\beta_{\mathrm{CMB}}) + D_\ell^{EB,\mathrm{CMB}} \sin(4\beta_{\mathrm{CMB}}), \label{eq:rot_bb} \\
    \tilde{D}_\ell^{EB,\mathrm{rot}} &= \frac{1}{2} \left(D_\ell^{EE,\mathrm{CMB}} - D_\ell^{BB,\mathrm{CMB}}\right) \sin(4\beta_{\mathrm{CMB}}) + {D_\ell^{EB,\mathrm{CMB}}\cos(4\beta_{\mathrm{CMB}})}. \label{eq:rot_eb}
\end{align}
\end{widetext}

Despite the degeneracy between the effective per-band rotation angle $\alpha_\nu$ and the isotropic cosmic birefringence angle $\beta_{\mathrm{CMB}}$, previous measurements have constrained $\beta_{\mathrm{CMB}} \lesssim 1^\circ$~\cite{BICEPKeck2025, Lee_2015_Birefringence}. Thus, the rotation impacts the $EB$ spectrum much more significantly than $EE$ or $BB$, since $EB$ is approximately linear in both $\beta_{\mathrm{CMB}}$ and $g$. Sample rotated spectra are illustrated in Fig.~\ref{fig:sample_rotate}.

When performing the multipole-dependent birefringence analysis, we replace the constant $\beta_{\mathrm{CMB}}$ in Eqs.~\eqref{eq:rot_ee}--\eqref{eq:rot_eb} with a piecewise-constant model defined by a Heaviside step function at a multipole breakpoint $\ell_b$:
\begin{equation}\label{eq:step_func}
    \beta_{\ell_b}(\ell) = \Delta\beta_{\ell_b} H(\ell - \ell_b),
\end{equation}
where $\Delta\beta_{\ell_b}$ characterizes the difference in rotation between low- and high-$\ell$ modes. The birefringence at low $\ell$ is not zero, but determined by the per-band rotation angles $\alpha_\nu$ such that $\Delta\beta_{\ell_b}$ isolates only the multipole-dependent deviation. We restrict our analysis to this step function formulation because it allows us to test for multipole dependence with the minimal increase in model complexity. While more elaborate functional forms for $\beta(\ell)$ could be employed, they introduce additional free parameters that increase the risk of overfitting; a comprehensive study of such models is reserved for future work.

\subsection{Galactic Dust and Synchrotron Treatment}
\label{app:dust_model}
Polarized galactic foregrounds are modeled for each auto or cross spectrum $XY$ between frequency bands $\nu_1$ and $\nu_2$ as a sum of dust and synchrotron components:

\begin{align}
D_\ell&^{XY,\mathrm{fg}}(\nu_1 \times \nu_2) = \nonumber \\
&f_d^{\nu_1}(\beta_d, BP^{\nu_1}) \cdot f_d^{\nu_2}(\beta_d, BP^{\nu_2}) \cdot A_d^{XY} \left( \frac{\ell}{\ell_0} \right)^{\alpha_d^{XY}} \nonumber \\
+ &f_s^{\nu_1}(\beta_s, BP^{\nu_1}) \cdot f_s^{\nu_2}(\beta_s, BP^{\nu_2}) \cdot A_s^{XY} \left( \frac{\ell}{\ell_0} \right)^{\alpha_s^{XY}},\label{eq:fg_eq}
\end{align}
where $\ell_0 = 80$ is the pivot multipole, and $BP^\nu$ denotes the bandpass response function for frequency $\nu$.

The coefficients $f_d$ and $f_s$ capture the scaling of dust and synchrotron power from the pivot frequencies to the actual bandpasses of the maps labeled $\nu_1$ and $\nu_2$ \cite{BK15}.

The foreground contribution is then added to the rotated CMB spectra:

\begin{align}
    \hat{D}_\ell^{XY}(\nu_1 \times \nu_2) &= \tilde{D}_\ell^{XY,\mathrm{rot}} + D_\ell^{XY,\mathrm{fg}}(\nu_1 \times \nu_2). \label{eq:fg_sum}
\end{align}
As detailed in Sec.~\ref{sec:data}, this analysis utilizes BICEP/Keck maps at 95, 150, and 220 GHz. At these frequencies, the synchrotron signal is negligible relative to thermal dust. We therefore fix the dust-synchrotron correlation parameter to zero, as the data lack the sensitivity required to meaningfully constrain this term~\cite{BICEPKeck2021b}.

\subsection{Per-Band Angle Rotation}
\label{app:det_rotation}
We apply a per-band polarization angle rotation $\alpha_\nu$ as an additional rotation on the spectra after including foreground contributions. This per-band rotation models the contributions from instrumental sources.
\begin{widetext}
\begin{align}
    \bar{D}_\ell^{EE}(\nu_1 \times \nu_2) &=
    \hat{D}_\ell^{EE}(\nu_1 \times \nu_2) \cos(2\alpha_{\nu_1}) \cos(2\alpha_{\nu_2}) + \hat{D}_\ell^{BB}(\nu_1 \times \nu_2) \sin(2\alpha_{\nu_1}) \sin(2\alpha_{\nu_2}) \nonumber \\
    &\quad - \hat{D}_\ell^{EB}(\nu_1 \times \nu_2) \cos(2\alpha_{\nu_1}) \sin(2\alpha_{\nu_2}) - \hat{D}_\ell^{BE}(\nu_1 \times \nu_2) \sin(2\alpha_{\nu_1}) \cos(2\alpha_{\nu_2}), \label{eq:anglecal_ee} \\
    \bar{D}_\ell^{BB}(\nu_1 \times \nu_2) &=
    \hat{D}_\ell^{EE}(\nu_1 \times \nu_2) \sin(2\alpha_{\nu_1}) \sin(2\alpha_{\nu_2}) + \hat{D}_\ell^{BB}(\nu_1 \times \nu_2) \cos(2\alpha_{\nu_1}) \cos(2\alpha_{\nu_2}) \nonumber \\
    &\quad + \hat{D}_\ell^{EB}(\nu_1 \times \nu_2) \sin(2\alpha_{\nu_1}) \cos(2\alpha_{\nu_2}) + \hat{D}_\ell^{BE}(\nu_1 \times \nu_2) \cos(2\alpha_{\nu_1}) \sin(2\alpha_{\nu_2}), \label{eq:anglecal_bb} \\
    \bar{D}_\ell^{EB}(\nu_1 \times \nu_2) &=
    \hat{D}_\ell^{EE}(\nu_1 \times \nu_2) \cos(2\alpha_{\nu_1}) \sin(2\alpha_{\nu_2}) - \hat{D}_\ell^{BB}(\nu_1 \times \nu_2) \sin(2\alpha_{\nu_1}) \cos(2\alpha_{\nu_2}) \nonumber \\
    &\quad + \hat{D}_\ell^{EB}(\nu_1 \times \nu_2) \cos(2\alpha_{\nu_1}) \cos(2\alpha_{\nu_2}) - \hat{D}_\ell^{BE}(\nu_1 \times \nu_2) \sin(2\alpha_{\nu_1}) \sin(2\alpha_{\nu_2}), \label{eq:anglecal_eb} \\
    \bar{D}_\ell^{BE}(\nu_1 \times \nu_2) &=
    \hat{D}_\ell^{EE}(\nu_1 \times \nu_2) \sin(2\alpha_{\nu_1}) \cos(2\alpha_{\nu_2}) - \hat{D}_\ell^{BB}(\nu_1 \times \nu_2) \cos(2\alpha_{\nu_1}) \sin(2\alpha_{\nu_2}) \nonumber \\
    &\quad - \hat{D}_\ell^{EB}(\nu_1 \times \nu_2) \sin(2\alpha_{\nu_1}) \sin(2\alpha_{\nu_2}) + \hat{D}_\ell^{BE}(\nu_1 \times \nu_2) \cos(2\alpha_{\nu_1}) \cos(2\alpha_{\nu_2}). \label{eq:anglecal_be}
\end{align}
\end{widetext}

\subsection{Band Power Window Functions}
\label{app:bpwf}
To directly compare theoretical model spectra with observational data, each theoretical auto or cross spectrum $\bar{D}_\ell^{XY}(\nu_1 \times \nu_2)$ is convolved with the corresponding band power window functions (BPWFs), computed as
\begin{equation}
    D_b^{XY}(\nu_1 \times \nu_2) = \sum_{\ell} W_{b\ell} \, \bar{D}_\ell^{XY}(\nu_1 \times \nu_2),
    \label{eq:bpwf_convolution}
\end{equation}
where $b$ indexes the band power and $W_{b\ell}$ is the weight assigned to multipole $\ell$. This convolution effectively bins the theory spectra into the same format as the data, incorporating deprojection, filtering, weighting, and instrumental filtering effects applied during data processing~\cite{BICEP2Keck2015}. For the EB cross spectrum, we approximate the window functions as the geometric mean of the EE and BB functions. While this approximation is known to introduce a multiplicative bias of a few percent~\cite{BICEPKeck2025}, this effect is negligible for our analysis given that the target birefringence signals are small and the uncertainty budget is dominated by statistical noise.

Each BPWF characterizes how power at multipole $\ell$ contributes to a given multipole band, typically peaking near the band center and tapering toward adjacent bands (Fig.~\ref{fig:bpwf_example}). While each band power receives contributions from a range of multipoles, the band edges, defined as the midpoint between adjacent band centers, encompass the range of dominant contributions. These band edges guide the choice of $\ell_b$ breakpoints used to capture multipole-dependent birefringence.

Figures~\ref{fig:after_bpwf} and \ref{fig:dust_bpwf} illustrate the effect of applying the BPWF convolution on EB spectra, highlighting changes for varying parameters \(g\) and dust foregrounds, respectively. Our analysis uses band powers corresponding to the multipole range $20 \lesssim \ell \lesssim 520$.

\subsection{Likelihood Calculation}
\label{app:likelihood_calc}
For all auto and cross spectra $XY$ and frequency pairs $(\nu_1, \nu_2)$, the binned band powers are concatenated to form the model prediction vector:
\begin{equation}
    \mathbf{D}^{\mathrm{model}}(\Theta) = \bigoplus_{XY, \nu_1, \nu_2} \left[ D_{b,1}^{XY}(\nu_1 \times \nu_2),\ldots, D_{b,{N_\mathrm{spec}}}^{XY}(\nu_1 \times \nu_2) \right],
    \label{eq:model_vector}
\end{equation}
where $N_\mathrm{spec}$ is the total number of $E$ and $B$ power spectra that are chosen to be included into the likelihood calculation. This stacking matches the structure of the observed data vector \(\mathbf{D}^{\mathrm{data}}\), enabling direct likelihood evaluation.
The Gaussian log-likelihood is then evaluated as
\begin{equation}
    -2 \ln \mathcal{L} =
    \left( \mathbf{D}^{\mathrm{model}}(\Theta) - \mathbf{D}^{\mathrm{data}} \right)^T
        \mathbf{\Sigma}^{-1}
    \left( \mathbf{D}^{\mathrm{model}}(\Theta) - \mathbf{D}^{\mathrm{data}} \right)
\end{equation}
where $\mathbf{\Sigma}$ is the band power covariance matrix. We use the same band power covariance matrix as previous analyses~\cite{BICEP2Keck2015},  where it is estimated using the baseline 499 simulations. To improve stability, terms with zero expected covariance are explicitly set to zero to reduce the noise from the finite number of simulated realizations. 

We use a Gaussian likelihood for computational efficiency and consistency with previous birefringence analyses. Given that the constraining power in our analysis focuses primarily on intermediate multipoles $\ell \sim 300\text{--}500$, the differences from the Hamimeche-Lewis likelihood are negligible for the parameter constraints presented here.

In summary, the sampled model parameter set is
\begin{equation}
\Theta = \left\{
\begin{aligned}
    &\beta_{\mathrm{CMB}},\quad A_{\mathrm{lens}},\quad g,\quad \{\alpha_\nu\}, \\
    &\{A_d^{XY}, \alpha_d^{XY}\},\quad \{A_s^{XY}, \alpha_s^{XY}\}, \\
    &\beta_s,\quad \beta_d
\end{aligned}
\right\},
\end{equation}
where
\begin{itemize}
    \item $\alpha_\nu$ are the per-frequency effective rotation angle parameters, used in Eqs.~\eqref{eq:anglecal_ee}--\eqref{eq:anglecal_be}.
    \item $\beta_{\mathrm{CMB}}$ is the isotropic cosmic birefringence angle. 
    \begin{itemize}
    \item In model-independent birefringence runs, $\beta_{\mathrm{CMB}}$ is modeled as a step function by substituting Eq.~\eqref{eq:step_func} into Eqs.~\eqref{eq:rot_ee}--\eqref{eq:rot_eb}.
    
    \item
    In EDE analyses, $\beta_{\mathrm{CMB}}$ is set to zero, so the $\alpha_\nu$'s become the overall per-band polarization rotation angle, whether from instrumentation or cosmological origin. 
    \end{itemize}
    
    \item $g$ is the photon-axion coupling constant that parametrizes parity-violating interactions in early dark energy models [Eq.~\eqref{eq:ede_eb}]. It is set to zero when EDE is not included.
    \item \(A_{\mathrm{lens}}\) controls the amplitude of lensing in the $BB$ spectrum. This parameter allows for variations in the lensing amplitude to account for uncertainties in the lensing measurement [Eq.~\eqref{eq:alens}].
    \item $A^{XY}$ and $\alpha_{XY}$ denote the amplitude and spectral tilt parameters for foreground components (dust and synchrotron) for each polarization spectrum type $XY \in \{EE, BB, EB\}$. The spectral indices $\beta_s$ and $\beta_d$ describe the frequency dependence of synchrotron and dust emission, respectively. Used in Eqs.~\eqref{eq:fg_eq}-\eqref{eq:fg_sum}.
\end{itemize}
In this analysis, we do not treat the EDE fraction $f_{\mathrm{EDE}}$ as a free parameter; attempting to simultaneously constrain $f_{\mathrm{EDE}}$ can introduce volume effects~\cite{Herold2022}. Furthermore, because the $EE$ power exceeds the $BB$ and $EB$ power by orders of magnitude, the $EE$ spectrum is effectively insensitive to the small rotations induced by birefringence. Consequently, we fix the background EDE parameters, which are primarily sensitive to $TT$, $EE$, and $TE$~\cite{Poulin_2019}, and instead adopt representative scenarios with $f_{\mathrm{EDE}}$ values motivated by prior literature: 0.012 (ACT), 0.07 (Planck), 0.087 (Planck + BOSS), and 0.127 (Planck + BOSS + SH0ES)~\cite{Eskilt2023, Murai2023, Herold2022, Calabrese2025} when fitting for the photon-axion coupling constant $g$.

We perform the Markov Chain Monte Carlo (MCMC) analysis using the \texttt{Cobaya} framework~\cite{Torrado2021}, imposing flat priors on all parameters with ranges chosen to be significantly wider than the expected posterior constraints. We verified that the posterior distributions for all sampled parameters are well contained within the prior ranges, with negligible probability density at the unphysical boundaries. Flexible configurations allow fixing subsets of parameters or data cuts (e.g., setting EB foregrounds to zero or excluding $EE$ and $BB$ spectra) to isolate specific effects. Convergence is assessed via the Gelman-Rubin statistic~\cite{Gelman1992} $R - 1 < 0.01$ (or $< 0.005$ when foreground amplitudes are fixed). These criteria follow the standards established by Dunkley \textit{et al.}~\cite{Dunkley2005}, ensuring that sampling errors in the posterior distribution are negligible compared to statistical uncertainties.

Parameter constraints are derived from marginalized posterior distributions. We marginalize over all $\alpha_\nu$ in our real results. Since this work primarily searches for a multipole-dependent effect, this marginalization will not affect the constraining power. We reserve the full discussion of the $EB$ spectra and their best fit $\alpha_\nu$ for future work.

\section{Data and Simulations}
\label{sec:data}
The data used in this analysis come from the \textit{Planck} Satellite~\cite{Planck2018III} and the BICEP/\textit{Keck} data through the 2018 observing season (BK18)~\cite{BICEPKeck2021b}. BK18 is a compilation of observations by the BICEP2, \textit{Keck} Array, and BICEP3 telescopes at the South Pole. The high-altitude, dry, and stable environment enables deep integration on small patches of the sky with minimal atmospheric interference, making it ideal for precise measurements of CMB polarization. Previous analysis of BK18 yielded a constraint on inflationary $B$-modes of $r<0.036$ at $95\%$ confidence~\cite{BICEPKeck2021b}.

\subsection{BK18 Dataset}
\label{sec:bk18}
The BK18 dataset includes observations from three generations of instruments:
\begin{itemize}
    \item \textbf{BICEP2}: A 150\,GHz receiver operated from 2010 to 2012, utilizing approximately 500 bolometers~\cite{BICEP2Keck2015}.
    \item \textbf{\textit{Keck} Array}: Operated from 2012 to 2019, initially with five 150\,GHz receivers, later expanded to include 95 and 220\,GHz receivers~\cite{BICEPKeck2021b}.
    \item \textbf{BICEP3}: A receiver observing at 95\,GHz, operational since 2016 with roughly 2500 bolometers~\cite{BICEPKeck2021b}.
\end{itemize}

The combined data cover an effective sky area of approximately 400 square degrees for BICEP2 and \textit{Keck} Array, and up to 600 square degrees for BICEP3. Maps employ an equirectangular pixelization with $0.25^\circ \times 0.25^\circ$ pixels. Observations target a region centered at RA = 0h, Dec = $-57.5^\circ$~\cite{BICEPKeck2025}.

Maps are constructed per observation phase after filtering time-ordered data. This process includes the removal of third-order polynomial trends, filtering of scan-synchronous signals, and removal of beam systematics by deprojection~\cite{BICEPKeck2021b}. Weighting is applied using inverse noise variance derived from the time streams, yielding estimates of the Stokes parameters $T, Q, U$ and their variances. The $Q$ and $U$ maps are converted to $E$ and $B$ using a map-based purification method described in BK-VII~\cite{bicep2keckvii}.

A suite of jackknife null tests is performed to validate data quality and identify systematic contamination~\cite{BICEP2Keck2015}. In this work, we use additional multipole band powers compared to the standard BK18 analysis to achieve stronger constraining power. We verify that these higher multipole band powers pass the standard suite of jackknife tests. Following the procedure described in Appendix B of BK-XIII~\cite{BICEPKeck2021b}, we repeat all $\chi^2$ and $\chi$ null tests using the same data-split definitions but extend the number of band powers from the baseline nine to fourteen, so that we use multipoles up to $\ell \approx 520$.

\begin{figure}
    \centering
    \includegraphics[width=1\linewidth]{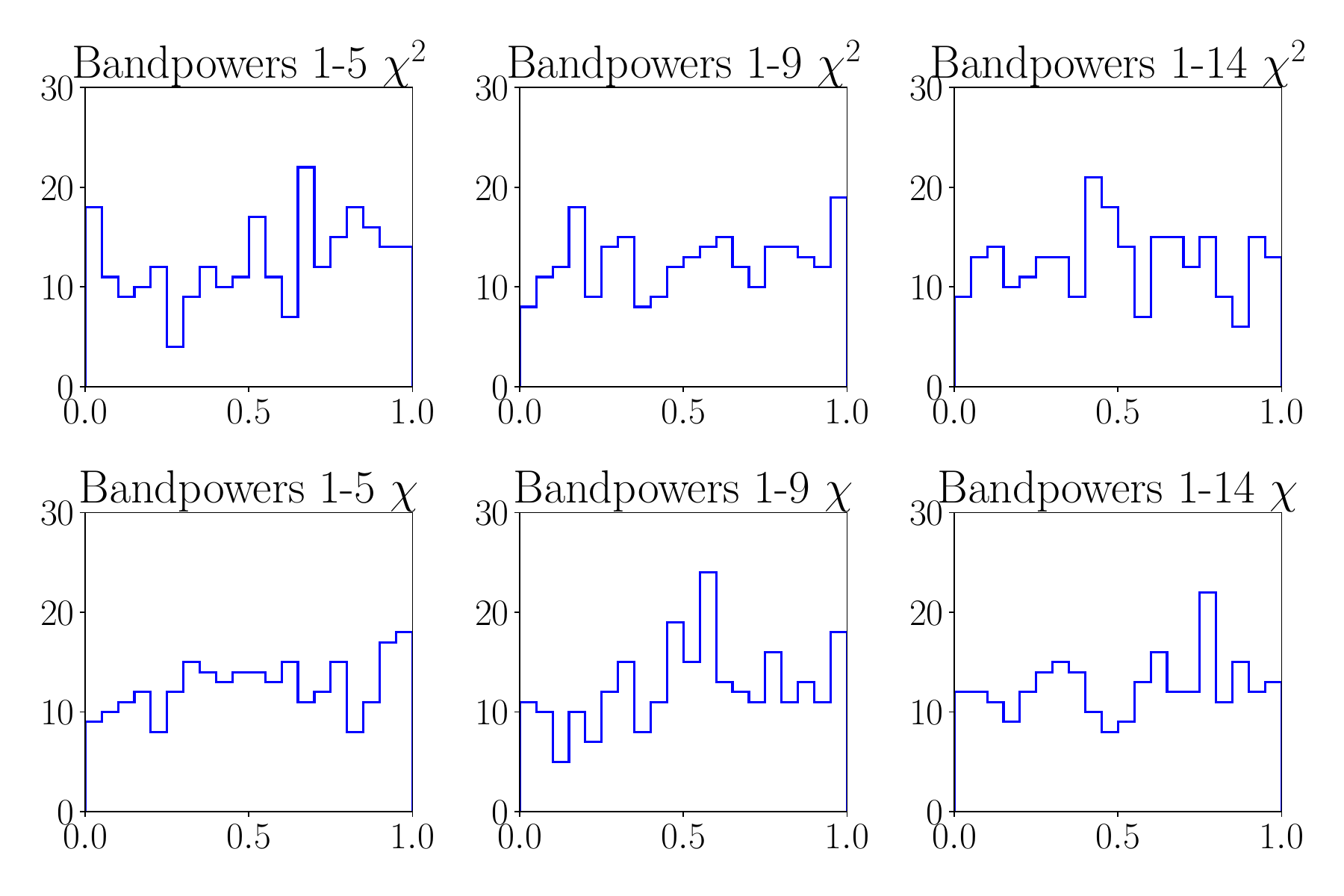}
    \caption{
        \textbf{Distributions of the jackknife $\chi^2$ and $\chi$ PTE values} for the Keck Array 2016, 2017, and 2018 220\,GHz data.  This figure is analogous to Fig.~12 of BKXIII~\cite{BICEPKeck2021b}, but with an additional rightmost column showing the extended multipole range (band powers 1--14), explicitly checking that the higher multipole band powers also satisfy the jackknife consistency criteria.
    }
    \label{fig:jackknife_220}
\end{figure}

We find that the distribution of probability to exceed (PTE) values remains consistent with expectations from simulations (Fig.~\ref{fig:jackknife_220}), with no increase in the number of extreme PTEs ($<0.01$ or $>0.99$) relative to the baseline analysis for any of the map frequencies.

We analyze the four mapsets of BK18: the large-field 95 GHz maps and three small-field maps at 95, 150 and 220 GHz. We also use external \textit{Planck} data, namely the PR4 maps at 30, 44, 143, 217 and 353 GHz~\cite{Planck2018III}, to constrain foreground parameters, particularly thermal dust, which dominates at higher frequencies. By marginalizing over foregrounds, we effectively separate their contribution from the cosmological signal.

\subsection{Simulations}
\label{sec:sims}
To characterize systematics, validate estimators, and compute uncertainties, we create realistic simulations of the observed data. We follow the standard procedure in the BICEP/\textit{Keck} power-spectrum analysis, with steps summarized below. See Refs.~\cite{bicep2keckvii, BICEP2014, BICEP2Keck2015} for further methodological details.

We construct two simulation sets, both using the standard BICEP/\textit{Keck} CMB and Gaussian dust modeling for simulating maps. The sets consist of 499 realizations processed through the full BICEP/\textit{Keck} power-spectrum-estimation pipeline. The only difference is the inclusion of an EDE-induced $EB$ signal in one of the two datasets:
\begin{itemize}
    \item \texttt{baseline}: These are the simulations used in the main BK18 analysis. They contain a lensed $\Lambda$CDM signal, sign-flip noise, and foreground dust, with no injected birefringence signal ($g = 0$).
    \item \texttt{EDE}: These are identical to the \texttt{baseline} simulations but with a $EB$ signal injected into the input spectra. This signal is generated by an early dark energy model with $f_{\mathrm{EDE}} = 0.07$ and $g = 1$ (Table~\ref{tab:ede_bestfits_reordered}). The parameter $f_{\mathrm{EDE}}$ represents the peak fraction of early dark energy during recombination, with the chosen value taken from the best-fit \textit{Planck} model in Murai \textit{et al.}~\cite{Murai2023}.
\end{itemize}

The simulation pipeline involves several steps. We first generate full-sky signal-only realizations of the CMB sky sampling from theoretical power spectra derived from \textit{Planck} best-fit parameters. We then beam convolve and reobserve the full-sky maps~\cite{bicep2keckvii}, encoding the real instrument’s scan strategy and time-domain filtering, to produce maps that reflect the actual sky coverage and resolution of the experiment.

Instrumental noise is added next. Noise realizations are derived from sign-flip simulations, created by randomly flipping the sign of real data subsets. This method preserves the statistical properties of the noise while suppressing the sky signal. We also include simulated foreground dust emission, which is modeled as Gaussian realizations drawn from an analytic power spectrum with parameters informed by \textit{Planck} observations. These realizations are scaled across frequencies using a modified blackbody spectrum.

After combining signal, noise, and dust, the maps undergo purification to reduce $E$-to-$B$ leakage~\cite{bicep2keckvii}. A purification matrix projects out ambiguous modes. Power spectra are then estimated using a ring-annulus Fourier estimator, which computes two-point correlations in annular bins in multipole space. This yields band power estimates for the $EE$, $BB$, and $EB$ spectra.

The power spectra are computed in 14 multipole band powers spanning $\ell = 20$ to 520, extending beyond the multipole range of $20\lesssim\ell\lesssim 355$ used in the $BB$ analysis to increase sensitivity to multipole-dependent birefringence signatures.

\section{Pipeline Validation}
\label{sec:simval}

We validate our two pipelines for constraining phenomenological multipole dependence and specific early dark energy models using realistic simulations introduced in Sec.~\ref{sec:sims}.

\subsection{Estimator for Recovering Multipole-Dependent Step-Function}
\label{app:ldiff_val}

We validate our estimator for multipole-dependent polarization rotation, modeled as a step function in $\beta(\ell)$, using our two simulation datasets: \texttt{baseline} and \texttt{EDE}. This approach introduces a single new parameter $\Delta\beta_{\ell_b}$, minimizing the total number of free parameters in our MCMC run. 

We fix the axion-photon coupling $g = 0$ and set all foreground amplitudes to zero so that all the multipole dependence is captured by the step function. We use only the $EB$ cross spectra in the data vector and fit for the four BK per-band polarization rotation angle parameters $\alpha_\nu$ and the step amplitude $\Delta\beta_{\ell_b}$.  The selected multipole breakpoints ($\ell_b = 265, 300, 335, 370, 405$) correspond approximately to the band edges near the second peak of the $EE$ spectrum. 

{ We restrict our analysis to the $\ell_b \in [265, 405]$ range because it represents the specific regime rigorously validated by our current simulations. Breakpoints in the off-peak regime below $\ell_b = 265$ lead to smaller signal amplitudes for the low-multipole bins, so the estimator becomes susceptible to uncharacterized systematics such as unconstrained dust foregrounds. Conversely, breakpoints above $\ell_b = 405$ reduce constraining power due to fewer high-$\ell$ band powers. The selected values strike a balance between statistical sensitivity and methodological robustness. }

For each $\ell_b$, we run MCMC chains on all 499 realizations in both simulation sets, using a convergence criterion of $R-1 < 0.03$. From each realization's marginalized 1D posterior, we extract the mean value of $\Delta\beta_{\ell_b}$ to serve as a single estimate. We then treat these 499 individual estimates as a new distribution to validate the estimator's performance. As shown in Table~\ref{tab:ldiff_peaks}, the ensemble average and 68\% credible interval of these 499 estimates are consistent with the null result in the \texttt{baseline} simulations; the recovered $\Delta\beta_{\ell_b}$ values are consistent with zero across all breakpoints, as expected. In contrast, for the \texttt{EDE} simulations, the distribution of recovered $\Delta\beta_{\ell_b}$ exhibits a $1-2\sigma$ level shift, consistent with the $\ell$-dependent signal predicted by the EDE model (Fig.~\ref{fig:step_function_demo}).

The mean values and $68\%$ credible intervals across all tested $\ell_b$ are summarized in Table~\ref{tab:ldiff_peaks} and plotted in Fig.~\ref{fig:ldiff_plot} for comparison. We find that the 68\% credible interval on $\Delta\beta_{\ell_b}$ is between $0.1^\circ$-$0.2^\circ$ over the angular scale to which we are sensitive. These validation results are within our expectations and demonstrate our sensitivity to multipole dependence.

\subsection{Estimator for Recovering EDE Photon-Axion Coupling}
\label{app:ede_val}
\begin{table*}[ht]
    \caption{\textbf{Best-fit cosmological parameters for EDE models} based on ACT and \textit{Planck} datasets, compiled from~\cite{Hill2022, HeroldThesis2023}. The ACT-motivated EDE0.012 benchmark corresponds to a minimal EDE fraction, while Planck-based fits span a wider range of $f_{\mathrm{EDE}}$ values up to 0.127. These parameters are used to generate the $EB$ spectra shown in Fig.~\ref{fig:ede_combined}.}
    \label{tab:ede_bestfits_reordered}
    \begin{ruledtabular}
    \begin{tabular}{lcccc}
        Parameter & ACT (EDE0.012) & \textit{Planck} (EDE0.07) & \textit{Planck}+BOSS (EDE0.087) & \textit{Planck}+BOSS+SH0ES (EDE0.127) \\
        \colrule
        $100\, \omega_b$          & 2.233  & 2.259  & 2.266  & 2.276 \\
        $\omega_{cdm}$            & 0.1198 & 0.1268 & 0.1285 & 0.1328 \\
        $100 \times \theta_s$     & 1.041  & 1.041  & 1.041  & 1.041 \\
        $\ln(10^{10} A_s)$        & 3.043  & 3.058  & 3.063  & 3.068 \\
        $n_s$                     & 0.965  & 0.9789 & 0.9833 & 0.9916 \\
        $\tau_{reio}$             & 0.054  & 0.0549 & 0.0563 & 0.0555 \\
        \colrule
        $f_{\mathrm{EDE}}$        & 0.012  & 0.070  & 0.087  & 0.127 \\
        $\log_{10} z_c$           & 3.500  & 3.563  & 3.559  & 3.562 \\
        $\theta_{i,\mathrm{scf}}$ & 2.749  & 2.742  & 2.747  & 2.761 \\
    \end{tabular}
    \end{ruledtabular}
\end{table*}
To assess the robustness of our inference pipeline in recovering the photon–axion coupling constant $g$, we carry out an extensive suite of validation tests using the two simulation datasets: \texttt{baseline} and \texttt{EDE}. We explore a range of EDE scenarios, foreground configurations, and injected per-band rotation signals. Our goal is to determine whether the pipeline can reliably recover the injected coupling parameter $g$ and the polarization rotation angles $\alpha_\nu$ for each frequency band.

As explained in Sec.~\ref{app:likelihood_calc}, we do not fit for the EDE fraction $f_{\mathrm{EDE}}$; instead, we choose a representative set of fiducial EDE parameters described in Table~\ref{tab:ede_bestfits_reordered}.

\subsubsection{Impact of Different Polarization Cross-Spectra}
Our MCMC likelihood supports fitting to different combinations of polarization spectra:
\begin{itemize}
    \item Marginalizing over foregrounds, free $A_s$ and $A_d$:
    \begin{itemize}
        \item $EE + EB$
        \item $EE + EB + BB$ ($A_{\mathrm{lens}}$ fixed to 1)
        \item $EE + EB + BB$ (free $A_{\mathrm{lens}}$)
    \end{itemize}
    \item Without marginalizing over foregrounds, $A_s$ and $A_d$ set to zero:
    \begin{itemize}
        \item $EB$ only
    \end{itemize}
\end{itemize}

For each test configuration, we run the pipeline on all 499 simulations, using the mean of each posterior as the best-fit value. We then compile the best-fit values from all realizations and compare them to the contours from a single realization; a comparable shape validates the predicted uncertainty (Fig.~\ref{fig:ede_g0}). We find unbiased recovery of the sampled parameters.

For the \texttt{baseline} simulation set (null birefringence), the recovered values of $g$ and $\alpha_\nu$ are consistent with zero across all tested configurations (Table~\ref{tab:g0_foregrounds}), demonstrating the accuracy and precision of the estimator in the absence of a birefringent signal. The $68\%$ credible interval on $g$ ranges from $0.31$ to $0.40$.

For the \texttt{EDE} set, using only $EB$ and $EE$ spectra yields unbiased recovery of both $g$ and $\alpha_\nu$ (Table~\ref{tab:g1_foregrounds}). However, when $BB$ spectra are included with $A_{\mathrm{lens}}$ fixed to 1, we observe a suppression in the recovered $g$ values and an increase of $\alpha_\nu$. The likelihood compensates for a $BB$ mismatch by shifting the effective per-band rotation angles $\alpha_\nu$. Due to parameter correlations, this inflation of $\alpha_\nu$ leads to a corresponding suppression in $g$.  This suppression suggests an inconsistency between the model and the simulation $BB$ lensing amplitude, which is also reflective of the existing uncertainties in the \textit{Planck} lensing models~\cite{Addison2024}. Allowing $A_{\mathrm{lens}}$ to float significantly mitigates this bias and restores parameter recovery to expected values.

\subsubsection{Impact of EDE Parameters Used in Generating the EB Curve}
We also examine how the recovered polarization rotation angles $\alpha_\nu$ and coupling $g$ vary as a function of the $EB$ template $D_\ell^{EB, \mathrm{EDE}}$ used in the likelihood, under both $g = 0$ and $g = 1$ simulation scenarios. We use a modified version of \texttt{CLASS}~\cite{Blas2011} developed by Murai \textit{et al.}~\cite{Murai2023} to generate four different $D_\ell^{EB, \mathrm{EDE}}$ curves using the EDE parameters described in Table~\ref{tab:ede_bestfits_reordered}. We then run the MCMC pipeline for each $EB$ template across all 499 simulations for both \texttt{baseline} and \texttt{EDE}. The \texttt{EDE} sims, including the injected $f_{\mathrm{EDE}}$ value, stay the same. Only the $EB$ cross spectra are included in this likelihood calculation. Since $EE$ and $BB$ are dominated by standard cosmology and foregrounds, this restriction allows us to isolate the dependence of our recovered parameters on the specific EDE template shape without introducing unnecessary covariance.

For the \texttt{baseline} simulation set (Table~\ref{tab:diffede_combined}, left), the recovered $g$ and $\alpha_\nu$ parameters are consistent with zero in all cases, demonstrating that the estimator remains unbiased regardless of the assumed EDE model used to construct $D_\ell^{EB, \mathrm{EDE}}$. However, for the case where $f_{\mathrm{EDE}} = 0.012$, the bias and uncertainty of $g$ increases. This occurs because the induced signal in $D_\ell^{EB, \mathrm{EDE}}$ is small for these parameters, making it more difficult to break the degeneracy between EDE-induced birefringence and effective per-band rotation.

In the \texttt{EDE} simulation set (Table~\ref{tab:diffede_combined}, right), where the signal is injected at $g = 1$, the recovered parameters show mild bias depending on the choice of $EB$ template. This is expected: the injected $D_\ell^{EB, \mathrm{injected}}$ in \texttt{EDE} corresponds to a specific set of EDE parameters where $f_{\mathrm{EDE}}=0.07$, while the $D_\ell^{EB, \mathrm{EDE}}$ used in the validation do not match for the remaining 3 of the 4 parameter sets. When the template does not match the injected signal, the likelihood fit can yield biased estimates.

This highlights the limitations of assuming a fixed functional form for the early dark energy signal. Since the true EDE parameters are not precisely known and are subject to change with future observations, this introduces systematic uncertainty into the inference of $g$. These results motivate the multipole-dependent approach described in Sec.~\ref{sec:angle_diff}, which allows for more flexible recovery of birefringence without assuming a specific EDE model. Nevertheless, we find that the $EB$-only analysis remains sensitive to birefringence effects at the tested coupling scale.

\subsubsection{Impact of Different Dust Models}
We run the MCMC pipeline on multiple datasets using a variety of dust models to assess the impact of foreground modeling on the estimator's recovery of $g$. The \texttt{baseline} dataset uses a Gaussian dust model for foregrounds in the BICEP/\textit{Keck} patch. To test robustness, we repeat the analysis using several alternative dust realizations. See Ref.~\cite{BICEPKeck2021b} for detailed descriptions of each model.

Across all cases, fitting jointly to $EE$, $BB$, and $EB$ spectra, the estimator consistently recovers values of $g$ consistent with zero (Table~\ref{tab:gdust_models}). The recovered values show minimal variation between models, indicating that foreground modeling does not significantly bias our constraints on $g$.

\subsubsection{Impact of Injected Per-Band Rotation}
To validate the robustness of our estimator against isotropic per-band rotation from instrumental sources, we inject known polarization rotation angles directly into the $EB$ spectra. This test evaluates whether the estimator can recover rotation angles correctly without biasing the inference of the photon-axion coupling parameter $g$.

We test three configurations of injected rotation angles across the four frequency bands (Table~\ref{tab:injected_rotation_combined}). For each configuration, we compute the corresponding theoretical $EB$ spectra, $\bar{D}_\ell^{EB,\ \text{injected}}(\nu_1 \times \nu_2)$, convolve them with the Band Power Window Functions (BPWFs), and add the resulting band powers to each realization in both the \texttt{baseline} and \texttt{EDE} datasets. Foregrounds and isotropic birefringence ($\beta_{\mathrm{CMB}}$) are set to zero in the likelihood model for this test. This simplification allows us to isolate the estimator's response to the injected rotation angles without introducing unnecessary covariance from foreground marginalization or cosmological parameter degeneracies.

The MCMC pipeline is then executed for each injected rotation scenario. The injected rotation angles are accurately recovered without introducing bias in the recovered $g$ (Table~\ref{tab:injected_rotation_combined}). Specifically, for both \texttt{baseline} and \texttt{EDE} simulation sets, the estimated $g$ remains consistent with the input value regardless of the rotation angle pattern. This test confirms that the estimator can distinguish between detector rotation-induced $EB$ and EDE-induced $EB$ and indicates that per-band rotation does not lead to artificial suppression or increase of the EDE signal.

\section{Results}

    \subsection{Constraints on Multipole-Dependent Birefringence}
    \label{sec:angle_diff}
    
At each $\ell_b$, we perform a MCMC analysis jointly fitting the four BK18 per-band rotation angles $\alpha_\nu$ and a single $\Delta\beta_{\ell_b}$. All chains satisfy the convergence criterion $R-1 < 0.005$.

\begin{figure}[ht]
    \centering
    \includegraphics[width=0.47\textwidth]{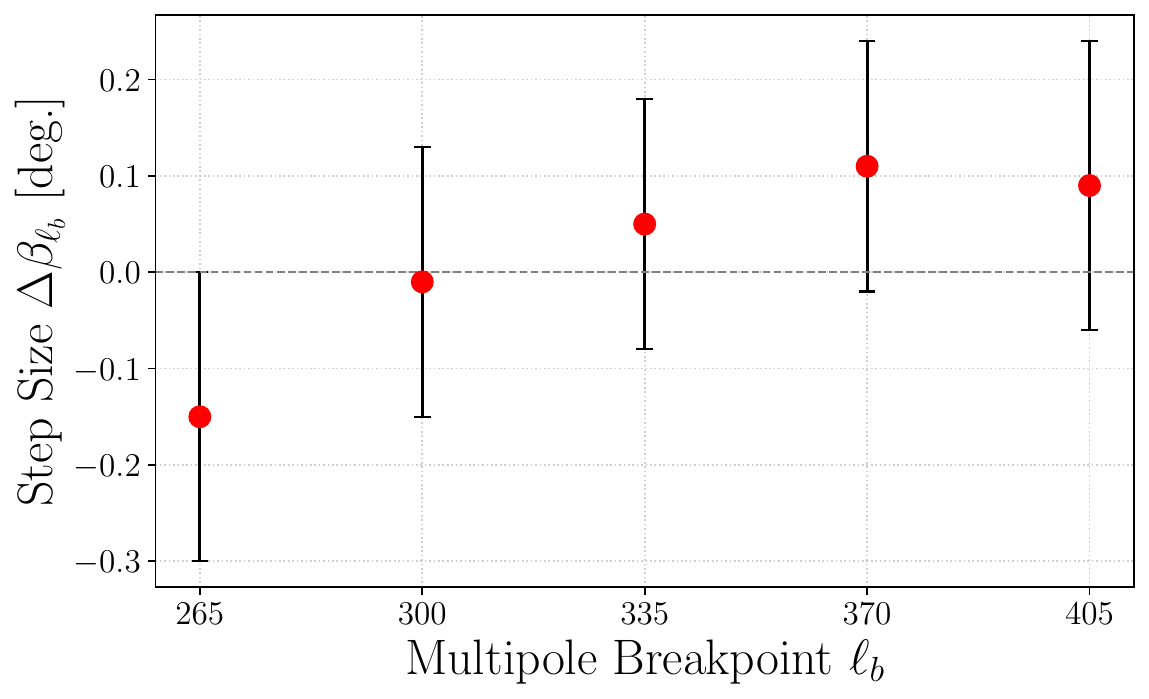}
    \caption{\textbf{Step Function Constraints.} Best-fit $\beta(\ell)$ step function for five values of the $\ell_b$ breakpoints with 1$\sigma$ uncertainty. All values are consistent with $\Delta\beta_{\ell_b}=0$.}
    \label{fig:eb_curves_lb}
\end{figure}

Fitting to the real BK18 data with the estimator validated in Sec.~\ref{app:ldiff_val}, Table~\ref{tab:angle_diff} and Fig.~\ref{fig:eb_curves_lb} show best-fit values for the $\beta(\ell)$ function at each $\ell_b$ breakpoint. In all cases, the data prefer values of $\Delta\beta_{\ell_b}$ consistent with zero at $\leq1\sigma$, and no statistically significant deviations are observed. 

\begin{table}[ht]
    \caption{\textbf{Step Function Constraints.} Posterior mean and 68\% credible intervals for the multipole-dependent rotation angle step $\Delta\beta_{\ell_b}$ at each multipole breakpoint. The step function model enables sensitivity to multipole-dependent rotation by splitting low- and high-$\ell$ band powers.}
    \label{tab:angle_diff}
    \begin{ruledtabular}
    \begin{tabular*}{\columnwidth}{cc}
        Breakpoint ($\ell_b$) & Size of Angle Difference ($\Delta \beta_{\ell_b}$) (deg) \\
        \colrule
        265 & $-0.15 \pm 0.15$ \\
        300 & $-0.01 \pm 0.14$ \\
        335 & $0.05 \pm 0.13$ \\
        370 & $0.11 \pm 0.13$ \\
        405 & $0.09 \pm 0.15$ \\
    \end{tabular*}
    \end{ruledtabular}
\end{table}

These results provide the first model-independent constraints on multipole-dependent birefringence from BICEP/\textit{Keck} data. They demonstrate the viability of CMB polarization spectra to probe time-varying parity-violating physics without strong assumptions about the underlying dynamics.

\subsection{Constraints on Early Dark Energy}
\label{sec:ede}\label{sec:ede_realdata}

We perform an extension to the analysis by Eskilt \textit{et al.}~\cite{Eskilt2023} using BICEP/\textit{Keck} data, as well as the \textit{Planck} spectra. In addition to $EB$-only fits, our analysis includes the $EE$ and $BB$ spectra and marginalizes over synchrotron and dust templates as described in Sec.~\ref{sec:isotropic_method}, allowing for more robust constraints that account for potential foreground contamination.

\begin{table*}[ht]
    \caption{\textbf{EDE Constraints.} Posterior mean and 68\% credible intervals on the photon-axion coupling \( g \) (units of $M_{\text{pl}}^{-1}$) at fixed EDE parameter values across multiple analysis configurations and data subsets.}
    \label{tab:ede_constraints}
    \begin{ruledtabular}
    \begin{tabular}{lcccc}
        Data in Likelihood & \( f_{\mathrm{EDE}} = 0.012 \) & 0.070 & 0.087 & 0.127 \\
        \colrule
        Eskilt-only & \(0.05 \pm 0.44\) & \(0.05 \pm 0.18\) & \(0.03 \pm 0.16\) & \(0.03 \pm 0.13\) \\
        BK18 EB-only & \(0.48 \pm 0.99\) & \(0.20 \pm 0.46\) & \(0.21 \pm 0.39\) & \(0.23 \pm 0.30\) \\
        BK18 EE+EB & \(0.41 \pm 1.01\) & \(0.12 \pm 0.41\) & \(0.14 \pm 0.38\) & \(0.17 \pm 0.31\) \\
        BK18 EE+EB+BB & \(0.28 \pm 1.01\) & \(0.07 \pm 0.40\) & \(0.11 \pm 0.37\) & \(0.11 \pm 0.31\) \\
        BK18 EE+EB+BB free $A_{\mathrm{lens}}$& \(0.21 \pm 1.02\) & \(0.06 \pm 0.43\) & \(0.08 \pm 0.37\) & \(0.10 \pm 0.31\) \\
    \end{tabular}
    \end{ruledtabular}
\end{table*}

We apply the same estimator validated in Sec.~\ref{app:ede_val} to real BK18 data, sampling over $g$ for each fixed value of $f_{\mathrm{EDE}}$.

\begin{figure}[h]
    \centering
    \includegraphics[width=0.75\columnwidth]{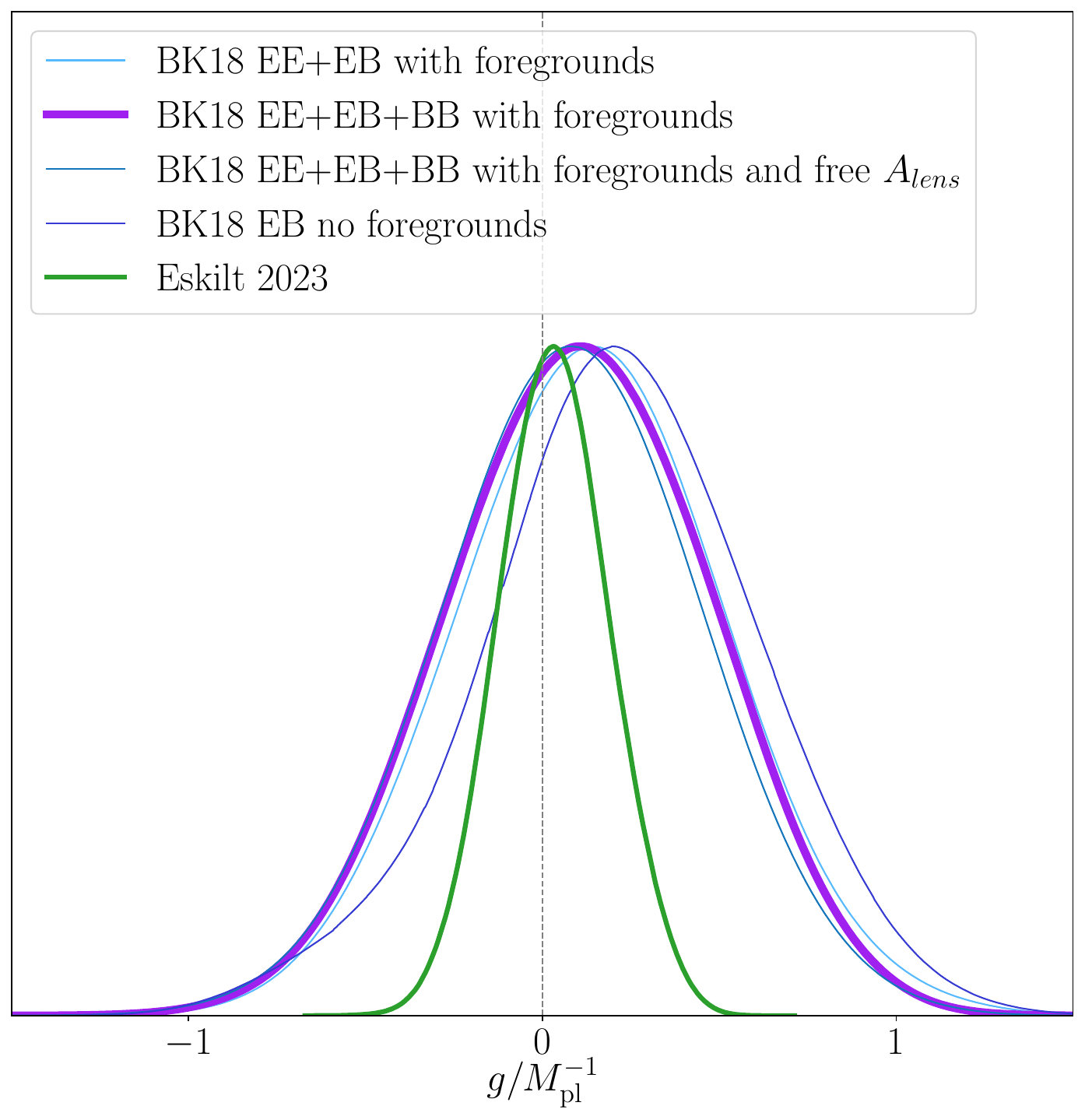}
    \caption{\textbf{EDE Constraints.} Posterior distribution on the axion-photon coupling \( g \) from BK18 (this work), overlaid with the constraint from \textit{Planck} EB analysis by Eskilt \textit{et al.}~\cite{Eskilt2023} with the EDE parameters from their baseline result.}
    \label{fig:g_constraints_overlay}
\end{figure}

Posterior distributions for $g$ are consistent with zero across all $f_{\mathrm{EDE}}$ values, as summarized in Table~\ref{tab:ede_constraints}. The result using Eskilt’s baseline EDE parameters is shown in Fig.~\ref{fig:g_constraints_overlay}. Marginalizing over foreground parameters does not significantly change the central values or widths of the posteriors, confirming the robustness of our constraints.

For the baseline EDE parameters ($f_{\mathrm{EDE}}=0.087$), Eskilt \textit{et al.}~\cite{Eskilt2023} reported a constraint of $g = 0.04 \pm 0.16$ using \textit{Planck} and BOSS data. For the same EDE parameters, our most robust configuration (EE+EB+BB) yields $g = 0.11 \pm 0.37$. While consistent with the \textit{Planck} result, our uncertainty is approximately a factor of 2 larger. This difference in our sensitivity reflects the trade-off between the full-sky coverage of \textit{Planck}, which accesses a wider range of multipoles, and the deep, focused integration of BICEP/\textit{Keck} on a smaller patch of sky.

Notably, the tightest constraints occur at higher $f_{\mathrm{EDE}}$, where the EDE-induced $EB$ templates deviate more significantly from a simple rotated $EE$ spectrum. The tighter constraints indicate stronger evidence against large signals of EDE, especially in multipole ranges sensitive to the time evolution of birefringence. Because of the limited sky coverage and angular resolution, the spectra from BICEP/\textit{Keck} data have a limited multipole range, limiting their statistical power compared to full-sky missions. 
\section{Conclusion}  
\label{sec:conclusion}

We have presented the first constraints on multipole-dependent cosmic birefringence using $EB$ power spectra from BICEP/\textit{Keck}. Parity-violating early dark energy models serve as a compelling theoretical test case, offering a physically motivated example of $\ell$-dependent effects in the early Universe.

To probe multipole-dependent effects, we introduced the phenomenological parameter $\Delta \beta_{\ell_b}$, describing a steplike change in rotation angle at multipole $\ell_b$. Across the tested range of breakpoints, $\Delta \beta_{\ell_b}$ was consistent with zero within uncertainties of $\sim 0.15^\circ$, placing phenomenological limits on a time-evolving parity-violating field during recombination.

We further constrained an axionlike EDE model with Chern-Simons coupling to photons, placing bounds on the axion-photon coupling $g$ for benchmark EDE parameters with varying energy fractions $f_{\mathrm{EDE}}$. Using BK18 data with multicomponent frequency coverage and robust foreground marginalization, we found the results to be consistent with the null hypothesis and to agree with existing limits from \textit{Planck}.

Importantly, this work establishes a versatile framework for testing multipole-dependent cosmic birefringence with CMB polarization, where more flexible parameterizations of $\beta(\ell)$ could be explored in future work. {Extended multipole-dependent simulations, specifically improving dust modeling and including full $EE$ and $BB$ spectra, will allow us to validate the estimator across a broader range of multipole breakpoints.}  While we avoided models requiring more than one free parameter for this work, modeling the rotation angle as a linear or periodic function of $\ell$ would allow for a wider class of multipole-dependent effects with only modest increases in model complexity. However, such extensions must be approached with care to avoid overfitting in the absence of strong theoretical priors. Incorporating foregrounds and other polarization spectra directly into the $\beta(\ell)$ analysis framework is another natural extension. 

Future calibration campaigns for BICEP3 are expected to yield precise measurements of absolute polarization angles, which are currently unconstrained in this analysis. As detailed in BICEP/\textit{Keck} XVIII~\cite{BICEPKeck2025}, establishing these absolute reference angles will break the degeneracy between isotropic cosmic birefringence and instrumental rotation. This advancement will enable direct constraints on $\beta_{\mathrm{CMB}}$ and significantly improve the separation of cosmological signals from instrumental systematics.

\section*{Acknowledgments}

The BICEP/\textit{Keck} projects have been made possible through a series of grants from the National Science Foundation, most recently including 2220444--2220448, 2216223, 1836010, and 1726917. The development of antenna-coupled detector technology was supported by the JPL Research and Technology Development Fund and by NASA Grants 06-ARPA206-0040, 10-SAT10-0017, 12-SAT12-0031, 14-SAT14-0009, and 16-SAT-16-0002. 

The development and testing of focal planes was supported by the Gordon and Betty Moore Foundation at Caltech. Readout electronics were supported by a Canada Foundation for Innovation grant to UBC. Support for quasi-optical filtering was provided by UK STFC grant ST/N000706/1. The computations in this paper were run on the Odyssey/Cannon cluster supported by the FAS Science Division Research Computing Group at Harvard University. The analysis effort at Stanford and SLAC is partially supported by the U.S. DOE Office of Science.

We thank the staff of the U.S. Antarctic Program and in particular the South Pole Station, without whose help this research would not have been possible. We extend special thanks to our heroic winter-overs during the observing seasons up until 2018: Robert Schwarz, Steffen Richter, Sam Harrison, Grantland Hall, and Hans Boenish. We thank all those who have contributed past efforts to the BICEP/\textit{Keck} series of experiments, including the BICEP1 team. We also thank the \textit{Planck} and WMAP teams for the use of their data. 

We are grateful to Johannes Eskilt, Kai Murai, Laura Herold, and Alessandro Greco for insightful discussions.

\clearpage
\appendix*
\onecolumngrid
\section{Supplementary Figures and Table}
\setcounter{figure}{0}    
\setcounter{table}{0}    
\renewcommand{\thefigure}{S\arabic{figure}}
\renewcommand{\thetable}{S\arabic{table}}

\begin{figure*}[ht]
    \centering
    \begin{subfigure}[h]{0.45\textwidth}
        \centering
        \includegraphics[width=\columnwidth]{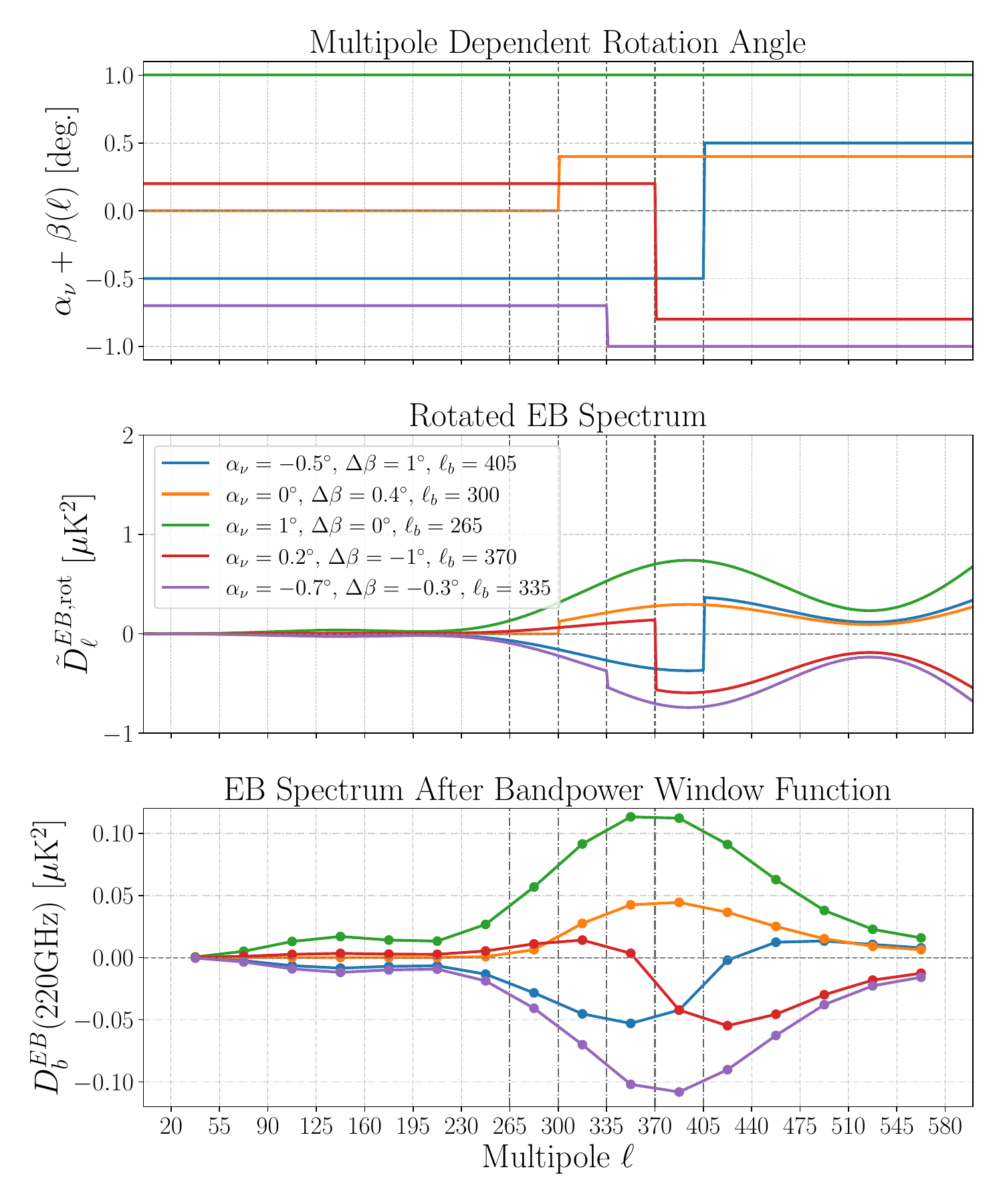}
        \caption{}
        \label{fig:lbreak_demo}
    \end{subfigure}
    \hfill
    \begin{subfigure}[h]{0.45\textwidth}
        \centering
        \includegraphics[width=\columnwidth]{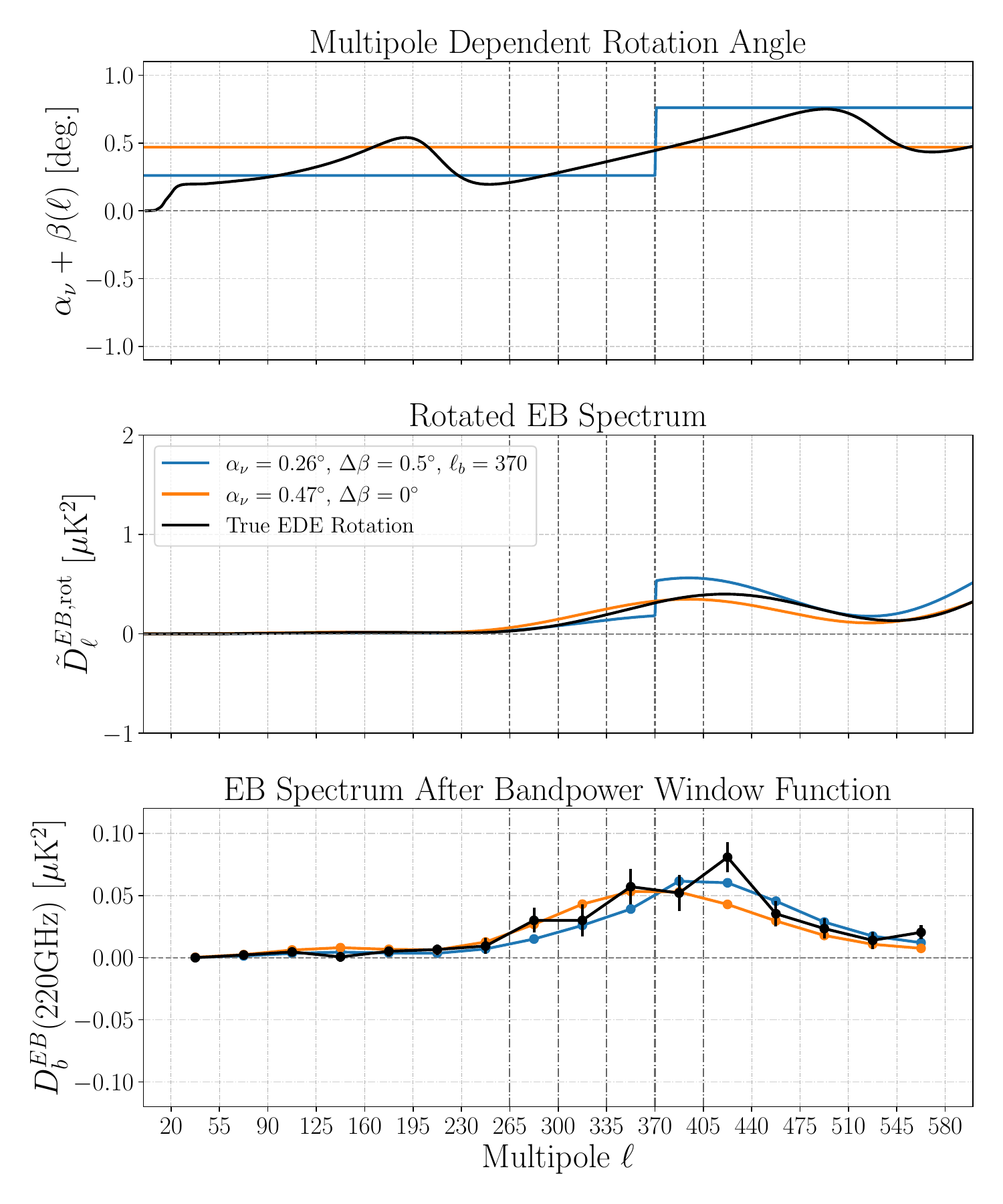}
        \caption{}
       
        \label{fig:ede_step}
    \end{subfigure}
    \caption{ \textbf{Visualizing step function modeling of $\ell$-dependent rotation.} (a)Examples of $\beta(\ell)$ modeled as step functions at various breakpoints $\ell_b$. Each curve represents a different step size $\Delta\beta_{\ell_b}$, demonstrating the effect of multipole-dependent changes in the polarization rotation angle. (b) Comparison between a step function $\beta(\ell)$ model (blue) and a constant-angle model (orange), fit to simulated \texttt{EDE} EB data, sim 79 (black). To produce a marginally better fit ($\chi^2 = 17.1$ vs $18.3$), the $\ell$-dependent estimator accommodates the features in the data by recovering a significant step size over the constant rotation model. In both cases, $g$ is set to zero so the EB spectrum is solely produced by this effective rotation.}
    \label{fig:step_function_demo}
\end{figure*}
    
\FloatBarrier

\begin{figure*}[ht]
    \centering
    \begin{subfigure}[b]{0.7\textwidth}
        \centering
        \includegraphics[width=\textwidth]{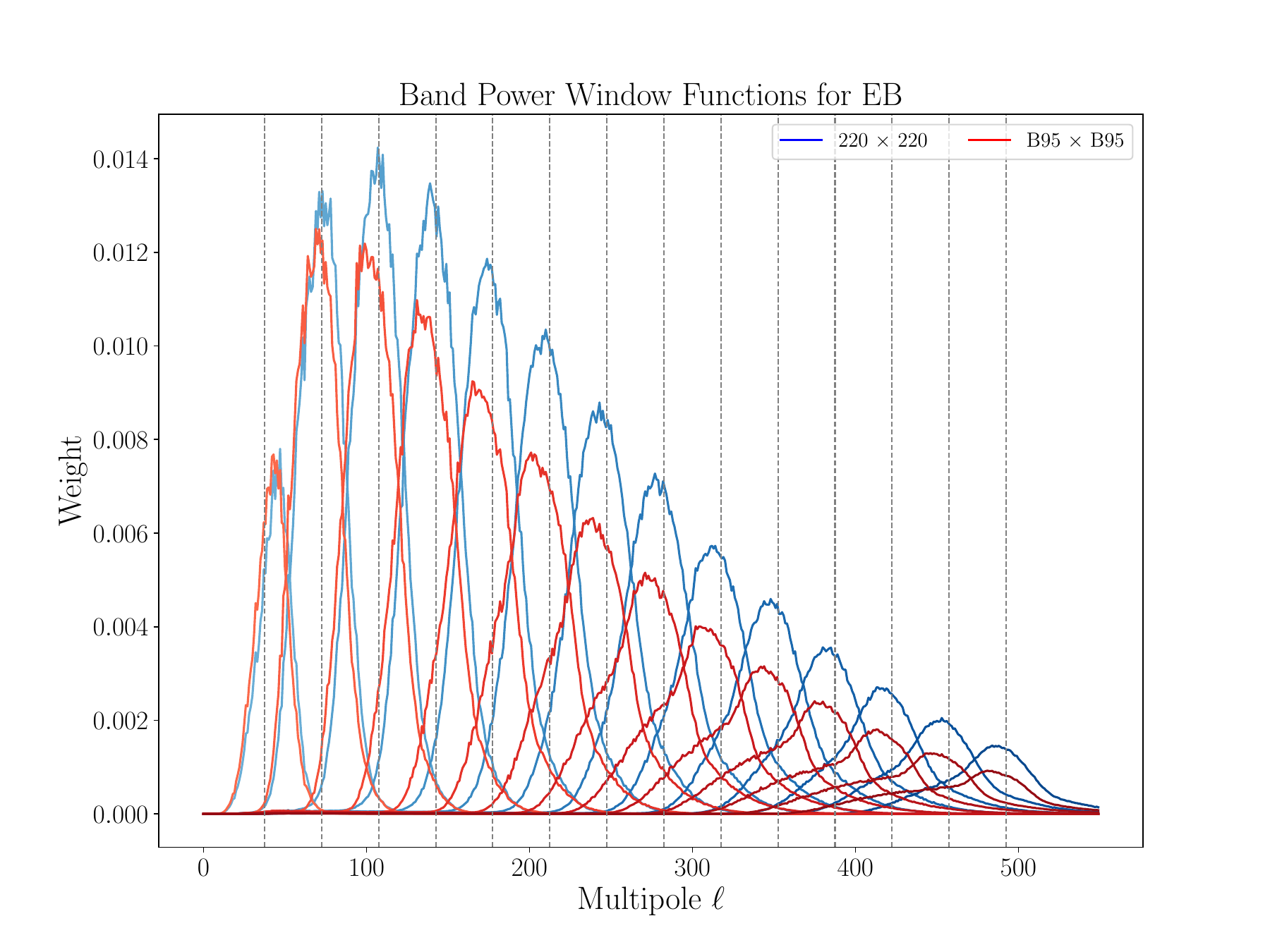}
        \caption{}
        \label{fig:bpwf_example}
    \end{subfigure}
    \hfill
    \begin{subfigure}[b]{0.25\textwidth}
        \centering
       \begin{ruledtabular}
    \begin{tabular}{cc}
        Bandpower \# & Nominal \(\ell\) Center \\
        \colrule
        1  & 37.5  \\
        2  & 72.5  \\
        3  & 107.5 \\
        4  & 142.5 \\
        5  & 177.5 \\
        6  & 212.5 \\
        7  & 247.5 \\
        8  & 282.5 \\
        9  & 317.5 \\
        10 & 352.5 \\
        11 & 387.5 \\
        12 & 422.5 \\
        13 & 457.5 \\
        14 & 492.5 \\
    \end{tabular}
    \end{ruledtabular}
        \caption{}
        \label{tab:bpwf_bins}
    \end{subfigure}

    \vspace{0.5cm} 
    \centering
    \begin{subfigure}[b]{0.47\textwidth}
        \includegraphics[width=\textwidth]{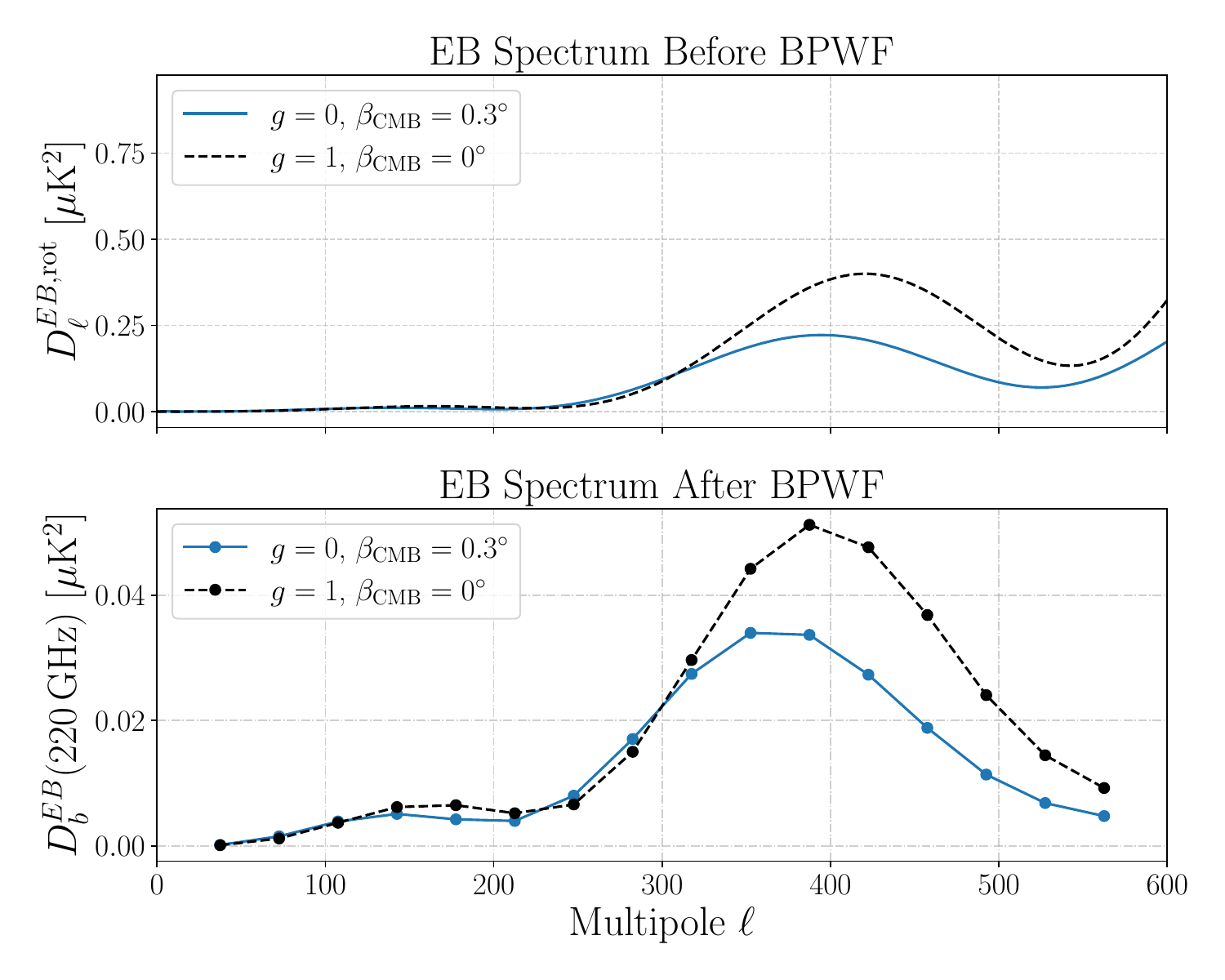}
        \caption{}
        \label{fig:after_bpwf}
    \end{subfigure}
    \hfill
    \begin{subfigure}[b]{0.47\textwidth}
        \includegraphics[width=\textwidth]{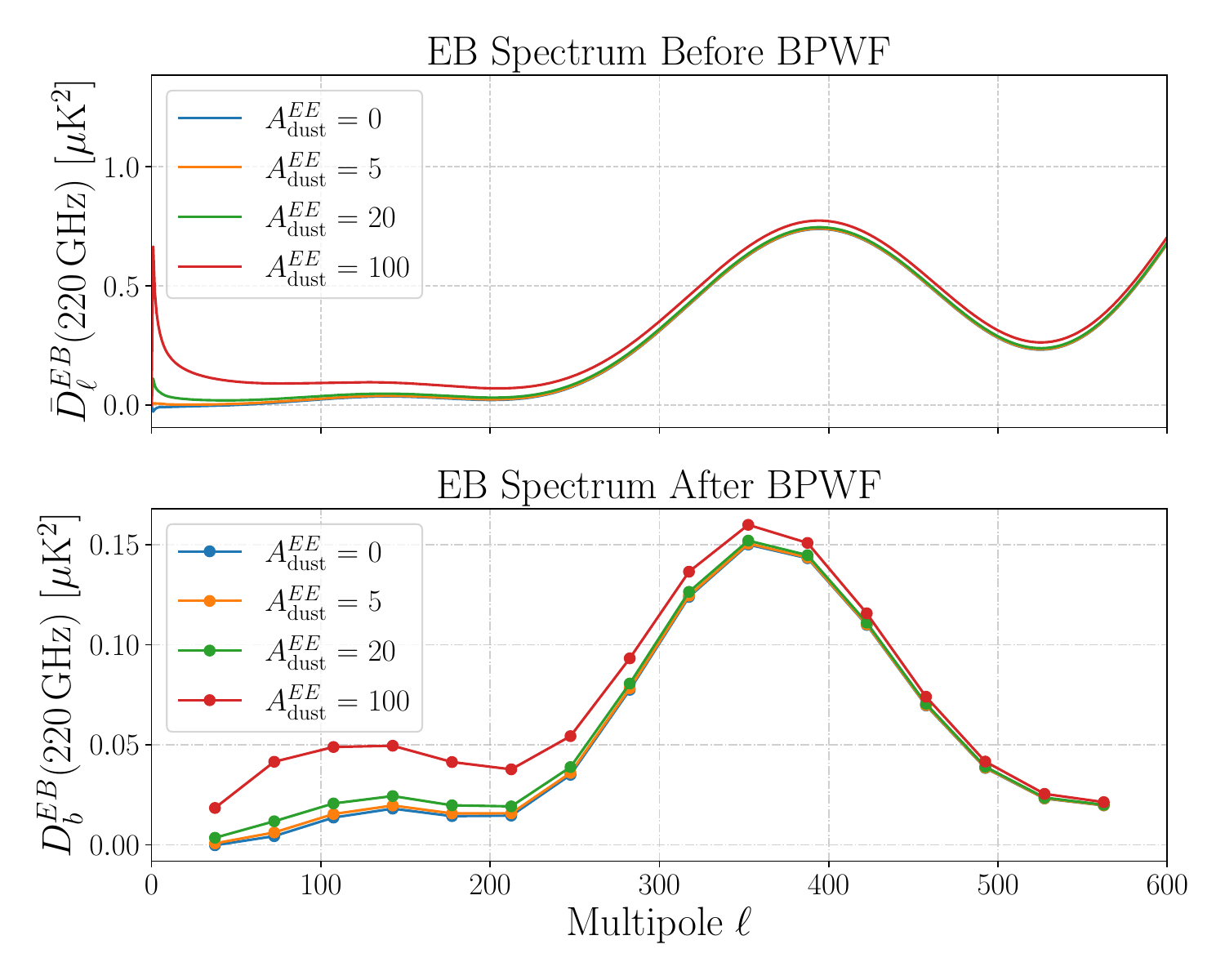}
        \caption{}
        \label{fig:dust_bpwf}
    \end{subfigure}
    \caption{\textbf{Visualizations of the Band Power Window Function and its effect on polarization spectra.} (a) Bandpower window functions (BPWFs) for selected cross spectra, showing contribution to each band power across \(\ell\). Dashed lines correspond to nominal band power centers. (b) Nominal band centers $\ell$ associated with each band power. The theoretical expectation values are calculated via the full convolution with the window functions shown in Fig.~\ref{fig:bpwf_example} according to Eq.~\eqref{eq:bpwf_convolution}. (c) Effect of BPWF on EB spectra for varying \(g\) and $\beta_{\mathrm{CMB}}$. No dust or per-band rotation applied. (d) Effect of BPWF on EB spectra with varying dust foreground parameters. Unspecified parameters are set to fiducial values (Table~\ref{tab:g0_foregrounds}). Due to suppression factors from instrumental and analysis filtering, the binned spectra in the lower panels appear at a reduced amplitude relative to the theory spectra.}
    \label{fig:bpwf}
\end{figure*}

\FloatBarrier

\begin{table}[h]
    \caption{\textbf{Validation of step function estimator.} The mean and 68\% credible interval of all recovered $\Delta\beta_{\ell_b}$ values (degrees) for each tested breakpoint $\ell_b$ over the 499 simulation realizations. These recovered values validate that the estimator can accurately constrain multipole dependence when none is present (null \texttt{$\Lambda$CDM} case) and detect multipole dependence when it is present in data (\texttt{EDE} case). The distribution of recovered $\Delta\beta_{\ell_b}$ means for $\ell_b = 370$ is shown in contour plot Fig.~\ref{fig:ldiff_triangle}.}\label{tab:ldiff_peaks}
    \begin{ruledtabular}
    \begin{tabular}{ccc}
        Stepfunction Breakpoint ($\ell_b$) & \texttt{baseline} ($g = 0$) & \texttt{EDE} ($g = 1$) \\
        \colrule
        265 & $0.01 \pm 0.16$ & $0.13 \pm 0.17$ \\
        300 & $0.01 \pm 0.14$ & $0.20 \pm 0.15$ \\
        335 & $0.02 \pm 0.12$ & $0.23 \pm 0.14$ \\
        370 & $0.01 \pm 0.14$ & $0.26 \pm 0.14$ \\
        405 & $0.01 \pm 0.17$ & $0.30 \pm 0.18$ \\
    \end{tabular}
    \end{ruledtabular}
\end{table}

\FloatBarrier  
\begin{figure*}[ht]
    \centering
    \begin{subfigure}[h]{0.48\linewidth}
        \includegraphics[width=\linewidth]{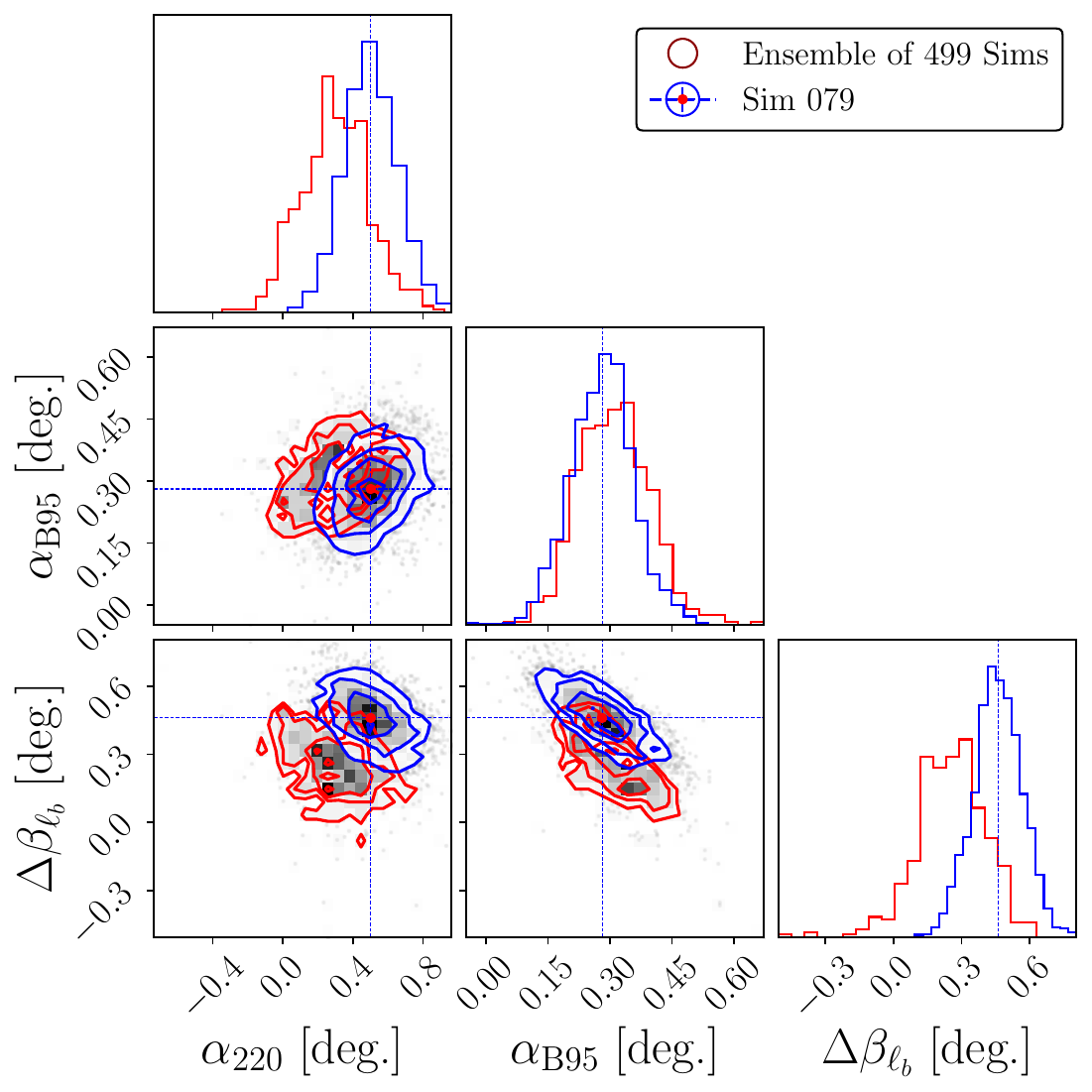}
        \caption{}\label{fig:ldiff_triangle}
    \end{subfigure}
    \hfill
    \begin{subfigure}[h]{0.48\linewidth}
        \centering
        \includegraphics[width=\linewidth]{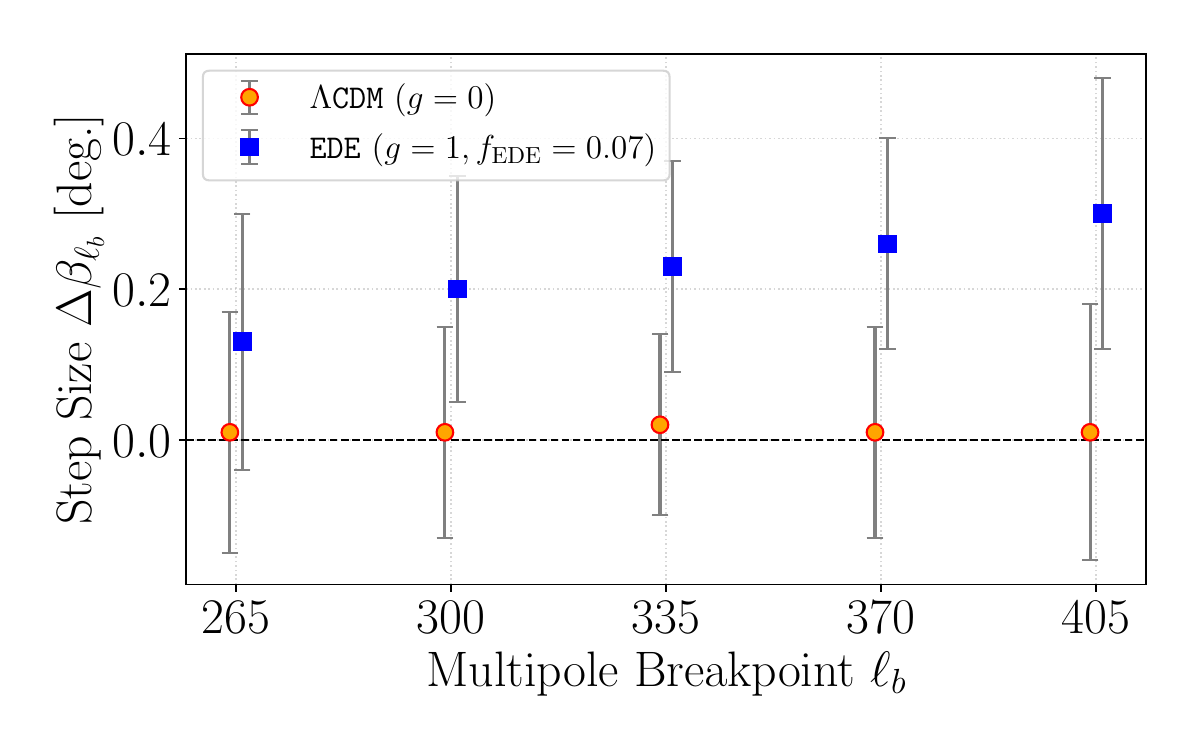}
        \caption{}\label{fig:ldiff_plot}
    \end{subfigure}
    \caption{\textbf{Validation of the model-independent step function estimator.} (a) Posterior distribution (where shaded regions represent 2D MCMC sample densities, and the contours represent the 68\%, 95\%, and 99\% credible intervals) of the step amplitude $\Delta\beta_{\ell_b}$ for a single \texttt{EDE} realization overlaid on the ensemble distribution of posterior means from 499 simulations (red). The blue dashed lines indicate the specific posterior mean for the representative realization (Sim 79), marking its position on the ensemble distribution with a red dot. (b) Recovered step amplitudes $\Delta \beta_{\ell_b}$ across all tested multipole breakpoints. Points represent the average of the 499 posterior means, while error bars represent the 68\% credible interval. The \texttt{baseline} set (blue) consistently recovers zero, while the \texttt{EDE} set (red) exhibits a nonzero, $\ell$-dependent signal characteristic of the injected birefringence.}
   
    \label{fig:ldiff_validation_combined}
\end{figure*}

\FloatBarrier

\begin{figure*}
    \centering
    \includegraphics[width=0.9\linewidth]{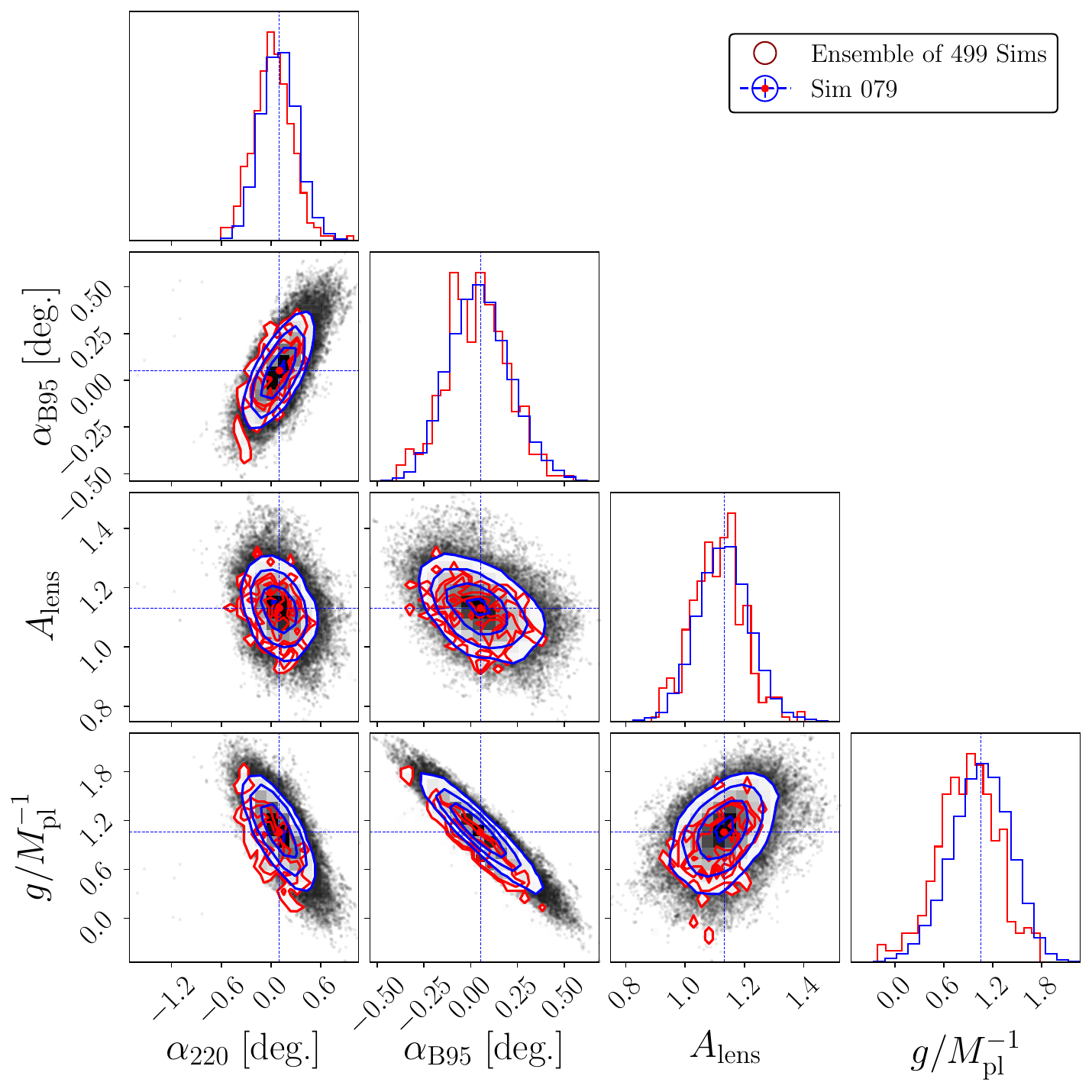}
    \caption{\textbf{Validation of parameter constraints for the EDE estimator.} Contours show the posterior from a single representative realization (blue) compared to the distribution of best-fit peaks from 499 simulations (red) in the \texttt{EDE} set ("All with scaled BB"). The agreement between the single-realization width and the ensemble scatter indicates that the likelihood accurately estimates parameter uncertainties. Blue dashed lines mark the mean of Sim 79 realization's posterior, and the red dot marks its position on the ensemble contour. For visualization, the posterior is projected onto select parameters of interest; all other foreground and instrument parameters (see Tables~\ref{tab:g0_foregrounds} and \ref{tab:g1_foregrounds}) are marginalized over.}
    \label{fig:ede_g0}
\end{figure*}
\FloatBarrier

\begin{table*}[ht]
    \caption{\textbf{Validation of parameter recovery on null (\texttt{baseline}) simulations.} Values represent the mean and 68\% credible interval of the ensemble of 499 individual posterior means recovered from the $\texttt{baseline}$ simulation sets. All recovered means are consistent with the fiducial input. The rotation angles are in units of degrees, $g$ is in units of $M_{\mathrm{pl}}^{-1}$, and dust amplitudes are in units of $\mu\mathrm{K}^2$. The spectral indices are unitless.}\label{tab:g0_foregrounds}
    \begin{ruledtabular}
    \begin{tabular}{lccccc}
        Parameter & All with free $A_{lens}$ & All & No BB & EB only & Fiducial\\
        \colrule
        $g$                     & $0.03 \pm 0.31$     & $0.03 \pm 0.35$ & $0.05 \pm 0.40$ & $0.03 \pm 0.40$ & 0 \\
        $\alpha_{220}$          & $-0.01 \pm 0.22$    & $-0.00 \pm 0.21$ & $-0.01 \pm 0.23$ & $-0.01 \pm 0.23$ & 0 \\
        $\alpha_{150}$          & $-0.02 \pm 0.17$    & $-0.01 \pm 0.17$ & $-0.02 \pm 0.18$ & $-0.02 \pm 0.18$ & 0 \\
        $\alpha_{K95}$          & $0.02 \pm 0.27$     & $0.00 \pm 0.24$ & $-0.01 \pm 0.25$ & $-0.00 \pm 0.25$ & 0 \\
        $\alpha_{B95lf}$        & $-0.02 \pm 0.15$    & $-0.02 \pm 0.16$ & $0.03 \pm 0.18$  & $-0.00 \pm 0.52$ & 0 \\
        $A_\mathrm{lens}$       & $1.05 \pm 0.08$     & ---             & ---             & ---             & 1 \\
        $A_\mathrm{dust}^{EE}$  & $7.7 \pm 1.3$       & $7.49 \pm 1.58$ & $8.05 \pm 2.07$ & --- & 7.5 \\
        $\alpha_\mathrm{dust}^{EE}$ & $-0.33 \pm 0.18$ & $-0.28 \pm 0.26$ & $-0.30 \pm 0.25$ & --- & -0.4 \\
        $A_\mathrm{dust}^{BB}$  & $3.86 \pm 0.8$      & $4.22 \pm 1.12$ & --- & --- & 3.75 \\
        $\alpha_\mathrm{dust}^{BB}$ & $-0.31 \pm 0.18$ & $-0.24 \pm 0.26$ & --- & --- & -0.4 \\
        $A_\mathrm{dust}^{EB}$  & $0.00 \pm 0.40$     & $0.00 \pm 0.41$ & $-0.01 \pm 0.46$ & $-0.01 \pm 0.49$ & 0 \\
        $\beta_\mathrm{dust}$   & $1.68 \pm 0.15$     & $1.68 \pm 0.16$ & $1.72 \pm 0.28$ & --- & 1.6 \\
    \end{tabular}
    \end{ruledtabular}
\end{table*}

\begin{table*}[ht]
    \caption{\textbf{Validation of parameter recovery on birefringent (\texttt{EDE}) simulations.} Values represent the mean and 68\% credible interval of the ensemble of 499 individual posterior means recovered from the $\texttt{EDE}$ simulation sets. The estimator recovers the injected signal across most configurations without bias. Selected parameters are shown in Fig.~\ref{fig:ede_g0}. The rotation angles are in units of degrees, $g$ is in units of $M_{\mathrm{pl}}^{-1}$, and dust amplitudes are in units of $\mu\mathrm{K}^2$. The spectral indices are unitless.}\label{tab:g1_foregrounds}
    \begin{ruledtabular}
    \begin{tabular}{lccccc}
        Parameter & All with free $A_{lens}$ & All & No BB & EB only & Fiducial\\
        \colrule
        $g$                     & $0.89 \pm 0.36$     & $0.71 \pm 0.32$ & $0.98 \pm 0.40$ & $1.00 \pm 0.40$ & 1 \\
        $\alpha_{220}$          & $0.02 \pm 0.22$     & $0.10 \pm 0.20$ & $-0.02 \pm 0.23$ &$-0.02 \pm 0.24$ & 0 \\
        $\alpha_{150}$          & $0.03 \pm 0.17$     & $0.12 \pm 0.15$ & $0.00 \pm 0.18$ & $-0.00 \pm 0.19$& 0 \\
        $\alpha_{K95}$          & $0.03 \pm 0.24$     & $0.11 \pm 0.23$ & $-0.01 \pm 0.25$ &$-0.01 \pm 0.26$ & 0 \\
        $\alpha_{B95lf}$        & $0.03 \pm 0.16$     & $0.12 \pm 0.14$ & $-0.02 \pm 0.18$ & $0.00 \pm 0.18$ & 0 \\
        $A_\mathrm{lens}$       & $1.11 \pm 0.09$     & ---             & ---             & ---             & 1 \\
        $A_\mathrm{dust}^{EE}$  & $7.88 \pm 1.45$     & $7.32 \pm 1.46$ & $7.93 \pm 1.98$ & ---             & 7.5 \\
        $\alpha_\mathrm{dust}^{EE}$ & $-0.38 \pm 0.21$ & $-0.31 \pm 0.25$ & $-0.34 \pm 0.24$ & ---             & -0.4 \\
        $A_\mathrm{dust}^{BB}$  & $3.84 \pm 0.84$     & $3.60 \pm 0.84$ & --- & ---             & 3.75 \\
        $\alpha_\mathrm{dust}^{BB}$ & $-0.37 \pm 0.24$ & $-0.28 \pm 0.25$ & --- & ---             & -0.4 \\
        $A_\mathrm{dust}^{EB}$  & $0.03 \pm 0.51$     & $0.01 \pm 0.38$ & $0.02 \pm 0.44$ & $0.02 \pm 0.48$ & 0 \\
        $\beta_\mathrm{dust}$   & $1.66 \pm 0.16$     & $1.61 \pm 0.16$ & $1.68 \pm 0.28$ & ---             & 1.6 \\
    \end{tabular}
    \end{ruledtabular}
\end{table*}

\FloatBarrier
\begin{table*}[ht]
    \caption{\textbf{Sensitivity of parameter recovery to the assumed EDE template shape.} Comparison of recovered angles $\alpha_\nu$ and coupling $g$ when using different initial theory templates ($f_{\mathrm{EDE}}$) for analysis across the 499 simulation realizations. In the null case (\texttt{baseline}, left), recovery is robust against template choice. In the signal case ($g=1$, right), slight biases appear if the analysis template differs from the true underlying signal, demonstrating the need for a model-independent estimator.}\label{tab:diffede_combined}
    \begin{ruledtabular}
    \begin{tabular}{lcccc c cccc}
        & \multicolumn{4}{c}{\texttt{baseline} $g=0$} && \multicolumn{4}{c}{\texttt{EDE} $g=1$} \\
        & \multicolumn{4}{c}{Template $f_{\mathrm{EDE}}$} && \multicolumn{4}{c}{Template $f_{\mathrm{EDE}}$} \\
        \cline{2-5} \cline{7-10}
        Parameter & 0.012 & 0.070 & 0.087 & 0.127 && 0.012 & 0.070 & 0.087 & 0.127 \\
        \colrule
        $\alpha_{220}$     & $-0.00 \pm 0.21$ & $-0.02 \pm 0.27$ & $-0.01 \pm 0.22$ & $0 \pm 0.17$    && $0.20 \pm 0.21$ & $0 \pm 0.30$   & $0.13 \pm 0.23$ & $0.34 \pm 0.19$ \\
        $\alpha_{150}$     & $-0.01 \pm 0.14$ & $-0.02 \pm 0.21$ & $-0.01 \pm 0.16$ & $0 \pm 0.10$    && $0.20 \pm 0.15$ & $0 \pm 0.23$   & $0.12 \pm 0.16$ & $0.35 \pm 0.10$ \\
        $\alpha_{K95}$     & $-0.02 \pm 0.25$ & $-0.03 \pm 0.32$ & $-0.04 \pm 0.26$ & $-0.01 \pm 0.23$ && $0.19 \pm 0.24$ & $0 \pm 0.29$   & $0.12 \pm 0.25$ & $0.35 \pm 0.23$ \\
        $\alpha_{B95lf}$   & $-0.01 \pm 0.13$ & $-0.03 \pm 0.22$ & $-0.01 \pm 0.15$ & $-0.01 \pm 0.07$ && $0.20 \pm 0.13$ & $0 \pm 0.23$   & $0.13 \pm 0.15$ & $0.36 \pm 0.07$ \\
        $g$                & $0.11 \pm 1.1$   & $0.02 \pm 0.46$  & $0.04 \pm 0.41$  & $0.03 \pm 0.35$ && $2.29 \pm 1.2$  & $0.97 \pm 0.45$ & $0.89 \pm 0.42$ & $0.75 \pm 0.37$ \\
    \end{tabular}
    \end{ruledtabular}
\end{table*}

\begin{table}[ht]
    \caption{\textbf{Systematics and robustness checks for dust.} Recovered $g$ values across 499 simulation realizations when analyzing null simulations generated with different non-Gaussian dust models; no significant bias is observed. Same labels as in Ref.~\cite{BICEPKeck2021b} are used for the alternate dust models.}
    \label{tab:gdust_models}
    \setlength{\tabcolsep}{3.5cm}
    \begin{ruledtabular}
    \begin{tabular}{lc}
        Dust Simulation Dataset & Recovered $g$ \\
        \colrule
        No Dust        & $0.04 \pm 0.31$ \\
        Gaussian Dust  & $0.03 \pm 0.31$ \\
        pysm1          & $-0.02 \pm 0.35$ \\
        pysm2          & $0 \pm 0.46$ \\
        pysm3          & $-0.03 \pm 0.38$ \\
        mhd            & $-0.07 \pm 0.33$ \\
        mkd            & $0.13 \pm 0.35$ \\
        gdecorr        & $0.04 \pm 0.32$ \\
        gampmod        & $0.03 \pm 0.32$ \\
        vansyngel      & $0.05 \pm 0.31$ \\
    \end{tabular}
    \end{ruledtabular}
\end{table}

\vspace*{\fill}

\begin{table*}[ht]
    \caption{\textbf{Robustness against injected detector rotation.} (a) For each configuration, the specific rotation angles (in degrees) injected into each frequency of all simulation realizations. (b) The mean and 68\% credible interval of the recovered coupling constant $g$ (in $M_{\mathrm{pl}}^{-1}$) for each configuration across 499 realizations. The consistency of the recovered values demonstrates that the estimator separates instrumental rotation from the EDE signal without bias.}\label{tab:injected_rotation_combined}
    
    \begin{subtable}{0.48\textwidth}
        \caption{Injected detector rotation}
        \label{tab:injected_rotation_a}
        \begin{ruledtabular}
        \begin{tabular}{lcccc}
            Configuration & 220 & 150 & K95 & B95lf \\
            \colrule
            Positive & 0 & 0.25 & 0.5 & 0.75 \\
            Negative & 0 & $-0.25$ & $-0.5$ & $-0.75$ \\
            Balanced & $-1$ & $-0.5$ & 0 & 0.5 \\
            Null injection & 0 & 0 & 0 & 0 \\
        \end{tabular}
        \end{ruledtabular}
    \end{subtable}
    \hfill
    \begin{subtable}{0.48\textwidth}
        \caption{Recovered coupling $g$}
        \label{tab:injected_rotation_b}
        \begin{ruledtabular}
        \begin{tabular}{lcc}
            Configuration & \texttt{baseline} & \texttt{EDE} \\
            \colrule
            Positive   & $0.04 \pm 0.37$ & $0.97 \pm 0.42$ \\
            Negative   & $0.03 \pm 0.38$ & $0.98 \pm 0.42$ \\
            Balanced   & $0.03 \pm 0.37$ & $0.97 \pm 0.42$ \\
            Null injection & $0.03 \pm 0.37$ & $0.97 \pm 0.41$\\
        \end{tabular}
        \end{ruledtabular}
    \end{subtable}
\end{table*}
\vspace*{\fill}
\clearpage
\twocolumngrid

\end{document}